\documentclass[reprint,amsmath,amssymb,txfonts,aps,pre,superscriptaddress,showpacs,floatfix]{revtex4-1} %,,prb
\usepackage{graphicx,bm,amssymb,amsmath,dcolumn,hyperref}
\usepackage{color,multirow}
\usepackage{braket}
\usepackage{mathrsfs}

\usepackage{float}
\usepackage{ulem}
\usepackage{natbib}
\begin{document}
\definecolor{red}{rgb}{1,0,0}
\newcommand{\red}[1]{\textcolor{red}{#1}}
\newcommand{\muast}{\mu^\ast}
\preprint{APS/123-QED}

\title{Charge density waves and superconductivity in the electron-positive fermion gas using a simple intuitive model. Part I: The model, instabilities, and phase diagram.}

%	\sout{Quantitative Electron-Electron Interaction in the Three-Dimensional Electron Gas Using Spin Local Field Factors from Variational Diagrammatic Quantum Monte Carlo Calculations}
% Elevation_609

\author{Carl A. Kukkonen} 
\email{kukkonen@cox.net}
\affiliation{33841 Mercator Isle, Dana Point California 92629, USA}

%	Autor 2
%	\author{Kun Chen}
%	\email{ kunchen@flatironinstitute.org }
%	\affiliation{Center for Computational Quantum Physics, Flatiron Institute, 162 Fifth Avenue, New York, NY 10010}

\date{\today}

\begin{abstract}

The electron-positive fermion gas in three dimensions and $T=0$ is modeled as two independent fermion gases interacting via the coulomb interaction. The main advantage of the simple model is that all existing results from the electron gas can be directly used for the positive fermion gas, which is the same as the electron gas, but scaled for the mass of the positive fermion. Additional screening from the positive fermions together with use of an accurate local field factor naturally introduces charge density waves in addition to the $q=0$ instability that occurs when the bulk modulus equals zero. The electron-positive fermion gas is completely specified by the density $r_s$ and the mass ratio $M/m$. Although the problem and model can be simply stated, the resulting phase diagram is complex and not fully understood. 

One simple result is that at the $q=0$ instability, the electron bulk modulus is positive, and the heavier positive fermion bulk modulus is negative; i.e., the electrons are stabilizing the heavier positive fermions. This contrasts with the model of the positive background stabilizing the electrons in conventional electron gas theory. A triple point that separates the uniform electron-positive fermion gas, an instability at $q=0$, and a charge density wave is obtained at a precise density when the mass ratio $M/m=4.97$. 

The results of the simple model are exact formulas, and are in close agreement with earlier numerical results obtained using density functional theory in their region of overlap. Using these results, the electron-electron, positive fermion-positive fermion and electron-positive fermion many body effective interactions are calculated in the following paper. For conditions close to the charge density wave, the positive fermion contribution to the electron-electron interaction, which is attractive and the source of the superconductivity, becomes large and significantly enhances the superconducting transition temperature, as well as leading to a large $T^2$ contribution to the normal state electrical resistivity.

\end{abstract}

%\pacs{Valid PACS appear here}% 
\maketitle

%%%%%%%%%%%%%%%%%%%%%%%%%%%%%%%%%%%%%%%%%%%%%%%%%%%%%%%%%%%%%%55 Introduction   %%%%%%%%%%%%%%%%%%%%%%%%%%%%%%%%%%%%%%%%%%%%%%%
\section{Introduction}

The electron–positive fermion gas has been theoretically investigated for more than 60 years \cite{ref1}. Early work focused on the electrons and holes in excited semiconductors and this problem was called the electron-hole liquid. In these systems, the masses of the electrons and holes are relatively close to each other (see Refs \cite{ref2,ref3,ref4} and references therein). At the opposite extreme, if the positive fermion is a proton with mass 1836 times the mass of the electron, the model was considered to possibly represent liquid metallic hydrogen \cite{ref5,ref6,ref7,ref8}. The degenerate electron-positive fermion gas can also be referred to as a plasma or liquid.

Most of the early theoretical investigations used center of mass coordinates which are convenient for two particle problems and provide easy comparison with exact solutions for bound states of excitons or the hydrogen atom. This can be solved exactly in the Hartree-Fock approximation and the focus was on calculating the correlation energy. The ground state energy was found to be close to the energy of an exciton, and the electron-hole gas was found to be unstable or metastable. Experimentally, however, the electron-hole gas does occur in optically pumped semiconductors.

Once the total energy is calculated as a function of density, the pressure and bulk modulus are obtained by differentiating the energy with respect to volume. The compressibility sum rule relates the response functions at small q to the bulk modulus that is the inverse of the compressibility. In prior work, linear response theory was used to calculate the response functions, phase diagram and effective interactions in the electron-hole gas. The effective electron-electron interaction was calculated using a two-fermion extension \cite{ref4,ref5,ref6,ref7,ref8,ref9} of the Kukkonen-Overhauser
interaction \cite{ref10} and the possibility of superconductivity was examined. The linear response theory and effective interactions are in the excellent and sophisticated textbook by Giuliani and Vignale \cite{ref11}.

The simple model of this paper does not use center of mass coordinates for the electron-positive fermion gas. It treats each gas independently and couples them by the coulomb interaction. Theory shows that the electron gas is uniform as long as the background is uniform. This observation is key to the simple approach of this paper. The solution for the positive fermion gas is the same as the electron gas, but with a different mass $M$. The positive fermions also have a uniform solution as long as the background is uniform. The positive fermions provide the neutralizing uniform background for the electrons and vice versa. When the bulk modulus of the combined system becomes zero, there is a real phase transition. Exchange and correlation amongst electrons and amongst positive fermions are included in the simple model. What is missing is electron-positive fermion correlation, which is neglected in this simple model.

The coulomb (or pseudopotential) interaction between the electrons and positive fermions is explicitly introduced in the calculation of the response functions. An important concrete observation is that an external test charge induces a density changes in both the electrons and positive fermions and both screen the external charge. The same additional screening also occurs if the positive charges were a lattice of ions. An external charge would slightly deform the lattice, which provides additional screening. This additional screening leads to the attractive potential between two electrons which is the basis for superconductivity. The additional screening from the positive fermions plays the role of phonons in BCS theory.

The additional screening from the positive fermions, when combined with accurate local field factors consistent with Quantum Monte Carlo calculations, naturally yields charge density waves for certain combinations of mass ratio $M/m$ and density $r_s$. Proximity to the charge density wave strongly enhances the attractive interaction between two electrons. This is explored in the following paper.

Early work used a local field factor based on the Hubbard approximation that did not predict charge density waves. More recently, density functional theory was used to calculate properties of the electron-positive fermion gas, which was shown to have instabilities including charge
density waves \cite{ref12}. The simple model of this paper yields a closed form of the response functions for all values of wave vector, density and mass ratio. The simple model results agree with density functional theory in their region of overlap, which includes the triple point, and
examples of charge density waves.

I believe that the simple intuitive model incorporates the much of the relevant physics and the neglected electron-positive fermion correlation energy will make only a small quantitative difference. This assumption can be checked by calculating the missing electron-positive fermion correlation energy as a function of density. The resulting phase diagram is complex and the natures of the new phases are unknown to me. The additional screening from the positive fermions is easy to understand intuitively. However the charge density waves also result from the precise value of the local field factor at intermediate wave vector and I have no intuition about this. An additional interesting feature is that the charge density waves can be extremely sensitive to density or mass ratio under certain circumstances.

I make no attempt to use the model to explain any experimental data, but there may be interesting applications. Just as in the electron gas, there is a unique density where the energy is a minimum for every mass ratio $M/m$. In using the electron gas results, the density parameter $r_s$ is
viewed as an adjustable parameter and taken as the local density in density functional theory. There may be a similar mapping process to use the results of the simple model. For mass ratios $M/m > 9$ the equilibrium density of the noninteracting simple model system is lower than the density for the onset of a charge density wave. Formally, the charge density wave negates the assumption of a uniform background and the model breaks down. However if the charge density wave has only a small amplitude, it may not affect the equilibrium density significantly and the overall electron-positive fermion gas has a charge density wave at equilibrium. Superconductivity would still be predicted and would coexist with charge density wave. There are many open questions, and I hope that this simple intuitive investigation will provoke further research.

The simple model is presented in detail in Appendix A and compared to previous work. For completeness, definition of the nomenclature and for the nonexperts, the response functions are re-derived in Appendices B \& C, including an additional local field factor $G_{12}$ that reflects the missing additional correlation. The numerical calculations in this paper using the simple model assume that $G_{12}$ is zero, but the response functions can be modified when this local field factor is eventually calculated. I consider $T=0$ where the electrons and positive fermions are each degenerate. However, there may be interesting physics for combinations of mass ratio and temperature where the heavier positive fermions are close to their Fermi temperature and the electrons are degenerate.

Section II presents results for the energy, pressure and bulk modulus of the simple model. Section III describes the $q=0$ instability which occurs when the system bulk modulus vanishes, the origin of the charge density waves, and different views of the complicated phase diagram of the electron-positive fermion gas as a function of wave vector $q/k_F$ , density represented by $r_s$ and the ratio of the positive fermion to electron masses $M/m$. The triple point at $M/m=4.97$ which separates the uniform gas, $q=0$ instability and the charge density wave regime is examined in detail. The charge densities introduced by a test charge are presented in Section IV. A summary and conclusions are given in Section IV.
\newcommand{\qtf}{q_{T\!F}}
%More recently, density functional theory was used to calculate properties of the electron-positive fermion gas, which was shown to have instabilities including charge density waves \cite{ref12}. The simple model yields a closed form of the response functions for all values of wave vector, density and mass ratio. The simple model results agree with density functional theory in their region of overlap, which includes the triple point, and examples of charge density waves.
%
%Section II presents results for the energy, pressure and bulk modulus of the simple model. Section III describes the $q=0$ instability when the system bulk modulus vanishes, the origin of the charge density waves, and different views of the complicated phase diagram of the electron-positive fermion gas as a function of wave vector $q/k_F$, density represented by $r_s$ and the ratio of the positive fermion to electron masses $M/m$. The triple point at $M/m=4.97$ which separates the uniform gas, $q=0$ instability and the charge density wave regime is examined in detail. A discussion and conclusions are given in Section IV.

%%%%%%%%%%%%%%%%%%	Section II %%%%%%%%%%%%%%%%%%%%%%%%%%%%%
\section{SIMPLE MODEL FOR THE ELECTRON–POSITIVE FERMION GAS}

The simple model used in this paper considers the electron-positive fermion gas as two independent systems coupled by the coulomb interaction. The electrons are considered to be in a uniform positive background. This problem has been studied extensively and the results are well known. Similarly, the positive fermions are considered in a uniform negative background, and the same electron gas results can be used directly because the Hamiltonian only depends on the charge squared, but the positive fermions can have a different mass $M$. The energies of these two independent systems are known, and the total system energy is the sum of the electron energy, positive fermion energy, plus an unknown additional correlation energy term.

The advantage of the simple model is that all of the results of the electron gas can be re-utilized. Exchange and correlation amongst electrons, and amongst the positive fermions are included, but the additional coulomb correlation between electrons and positive fermions is not. 

The electron gas has a uniform solution as long as the background is uniform. The same is true for the positive fermion gas. The simple model assumes that the positive fermions are the uniform background for the electrons, and vice versa. The model breaks down when either density becomes nonuniform. When the compressibility of the total system becomes negative at $q=0$, the system is truly unstable. If the instability occurs at finite $q$, this signals a charge density wave.

Calculation of the additional electron-positive fermion correlation energy that is missing from this simple model is beyond the capabilities of the author, and would represent completion of the quantitative description of the electron-positive fermion gas.  

In this paper, the role of the additional correlation in the response functions is shown, but in the numerical calculations, it is neglected. In Appendix A, the simple model is compared to earlier calculations, and it is argued that the additional correlation energy is relatively small and independent of density near the energy minimum. This means that the energy minimum, pressure and bulk modulus of the simple model are in close agreement with previous work.  I believe that the simple model contains much of the relevant physics and that the additional correlation energy will only provide small corrections to the predicted densities of the energy minimum, compressibility divergence, phase diagram and wave vectors of the charge density waves.

The assumption that the additional correlation energy is small and independent of density near the energy minimum is a crucial assumption that may not hold over the full range of densities and mass ratios.

The details of the simple model are given in Appendix A. The total energy is the sum of the electron energy and the positive fermion energy, and similarly for the pressure and bulk modulus. A heavier positive fermion has a lower minimum energy at a smaller $r_s$ compared to the electron. The combined energy has a unique minimum at an $r_s$ between that of the energy minimum of the electrons and the positive fermions. At the $r_s$ of the combined energy minimum, the pressure of the electron gas is positive, and the pressure of the positive fermion gas is equal, but negative, and the total pressure is zero. The bulk modulus of the electron-positive fermion gas is found to be zero at an $r_s$ that is $1.25$ times the $r_s$ of the energy minimum independent of the ratio of the masses.

%%%%%%%%%%%%%%%%%%%%%% Figure 1 %%%%%%%%%%%%%%%%%%%%%%%%%%%%%%%%%%%%%%%
\begin{figure}[h!]
	\centering
    \includegraphics[width=1.0\columnwidth]{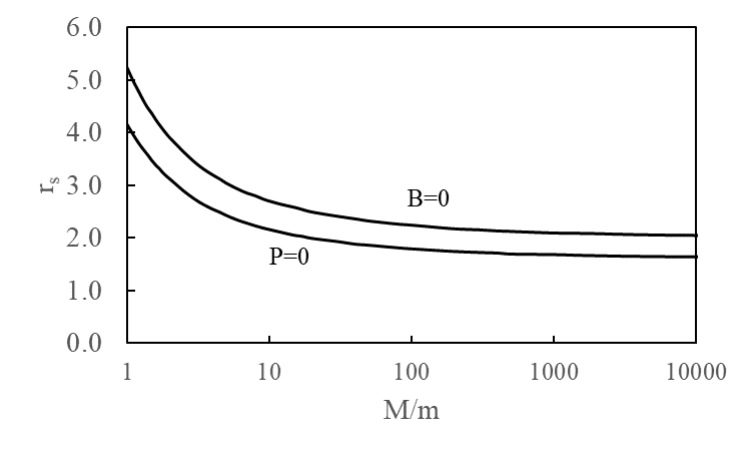}
    \caption{Linear measure of density $r_s$ where the energy of the electron-positive fermion in the simple model gas is minimum, and the $r_s$ where the bulk modulus is zero and the system is unstable as a function of the ratio of the positive fermion mass to the electron mass, $M/m$.}
    \label{fig1}
	\end{figure} 
%%%%%%%%%%%%%%%%%%%%%%%%%%%%%%%%%%%%%%%%%%%%%%%%%%%%%%%%%%%%%%%%%%%%%%%

Fig. \ref{fig1} clearly shows that there is an asymptotic minimum value of $r_s$ for the energy minimum near $1.62$ as the positive fermion mass goes to infinity. This occurs because the kinetic energy of the heavier particle goes to zero. Interestingly, this is near the value of $r_s$ estimated for metallic hydrogen \cite{ref13}. There is also a maximum $r_s$ of $4.19$ when the two masses are equal. This is the same as for the electron gas in a uniform background.

An interesting observation from Appendix A is near the instability, the bulk modulus of the electrons is positive and the bulk modulus of the positive fermions is negative. The lighter electrons are stabilizing the system. In the uniform electron gas, it is assumed that the uniform background is providing stability so that one may not be concerned when the electron gas bulk modulus becomes negative.

Just as in the uniform electron gas, for each mass ratio there is a unique energy minimum for point fermions. To achieve a different density, one would need to apply pressure to reduce $r_s$ or add a pseudo potential as done in the electron gas \cite{ref18} to achieve a larger $r_s$.

The simple model is physically intuitive and allows mental pictures of how the electrons and positive fermions are responding. The largest quantitative benefit of the simple model is that the bulk modulus and susceptibility properties are known for all densities because these quantities are known for the uniform electron gas. These are crucial for the calculations of the response functions. Note that the effective density $r_s$ of the positive fermions is $M/m$ times the density of the electrons. For large mass ratios, the effective $r_s$ of the positive fermions becomes very large. At this low density, the Wigner lattice has a very slightly lower energy than the uniform solution. The dependence of the energy of the Wigner lattice is similar to that of the uniform electron gas \cite{ref11}.  The response functions should be very similar. 

In Appendix A, the simple model used in this paper is shown to compare favorably to previous work on the electron-hole liquid with mass ratio $M/m=1-4$ and to calculations on the electron-proton gas as a model for metallic hydrogen with $M/m=1836$. The comparison indicates that the simple model may be valid for both extremes of mass ratio.

%%%%%%%%%%%%%%%%%%%%%%%%%%%%%%%%%%%      Section III     %%%%%%%%%%%%%%%%%%%%%%%%%%%%%%%%%%%%%%%%%%%%%%%%%%%%%    
\section{STABILITY, CHARGE DENSITY WAVES AND PHASE DIAGRAM}
%%%%%%%%%%%%%%%%%%%%%%%%%%%%%%%%%%%%%%%%%%%%%%%%%%%%%%%%%%%%%%%%%
%The general linear response theory of the two-fermion gas was calculated by Vashishta et al.\cite{ref3} in $1974$. The theory was extended to include the electron-electron interaction in 1985 in Ref. \cite{ref4}. For completeness and easy access for the nonexpert reader these quantities are re-derived using the intuitive Kukkonen-Overhauser \cite{ref10} approach in Appendix B. 
Instabilities are indicated by divergences in the linear response functions which determine the system response to an external potential. The instabilities found in this paper are at $q=0$ and for mass ratios $M/m \geq 4.97$ charge density waves at finite $q$. The $q=0$ instability occurs at the point where the bulk modulus of the system becomes zero. Charge density waves do not occur in the single fermion gas, but do occur in the electron-positive fermion gas. The charge density waves are a result of the additional screening from the positive fermions, and are a sensitive function of the local field factor at finite wave vector near $q = 1-2 \, k_F$. The wave vector dependence of the local field factors of the electron gas was determined by Quantum Monte Carlo calculations using supercomputers (see Refs. \cite{ref17,ref20,ref23} and references therein). This data was not available earlier and the then popular Hubbard approximation does not yield charge density waves \cite{ref4}. 

The calculation of the response functions and effective interactions in the electron-positive fermion gas in terms of local field factors is straightforward but somewhat tedious. In order to understand the two fermion system, it is necessary to understand the single fermion gas in a uniform background. A typical reader is likely not an expert in this field and the following paragraphs may help before jumping straight into the analysis.

Fermions have two spins and the response of fermions with spin up may be different than that for spin down, depending on the type of external potential. If the external potential is a point charge, it affects spin up and spin down fermions in the same way and the total induced density is simply the sum of the induced densities of fermions with spin up and spin down. If the external potential is a magnetic field, the fermions with spin up have a different response than those with spin down. The total induced magnetization is proportional to the difference between the spin up and spin down induced densities. If the external potential is a fermion with spin up, it has a different interaction with other fermions of spin up from the interaction with fermions with spin down. This is well known and explained in textbooks such as Ref. \cite{ref11}. For completeness, easy access for the non-expert reader and to define the notation used in this paper, this linear response theory for a single type of fermions is re-derived in Appendix B using the intuitive Kukkonen-Overhauser approach.

For the electron-positive fermion gas, the same approach holds except now there are two different types of fermions. The linear response theory for this system was originally presented in Ref. \cite{ref3} in 1974 and extended to include the electron-electron interaction in Ref, \cite{ref4} in 1985. These results are re-derived in Appendix C. The nonexpert reader is highly encouraged to look at these two appendices to see the spirit of the calculations and the source of the notation. 

The total density response is indicated with a plus subscript $\Delta n_+ = \Delta n_\uparrow + \Delta n_\downarrow$, $ V_+^{\text{ext}} = (V_\uparrow^{\text{ext}} + V_\downarrow^{\text{ext}})/2$ and $V_+^{\text{eff}} = (V_\uparrow^{\text{eff}} + V_\downarrow^{\text{eff}})/2$ with corresponding notation for the differences. With two types of fermions, the additional subscript $1$ refers to the positive fermions and $2$ indicates the electrons.

The effective interactions calculated in Appendix B do not assume that the bare interactions between electrons and positive fermions, and positive fermions with each other are purely coulomb interactions, which would allow a pseudo-potential to be utilized. In this section, however, these bare interactions are all assumed to be coulombic. The bare interaction between like particles is repulsive and has a positive sign.% The bare interaction between different particles is attractive and has a negative sign. The subscript 1 refers to the positive fermions, and the subscript 2 is for electrons. The $+$ sign indicates the total density response which is the sum of the response of the spin up and spin down particles. The $-$ sign indicates the magnetic response which is proportional to the difference in response between spin up and spin down particles.
 The following equations are derived in Appendix C and the equations there have the numbers (\ref{eq_c15}-\ref{eq_c16}). It is assumed that $V_{11}^b = v$ and $V_{12}^b = -v$. 
%%%%%%%%%%%%%%%%%%%%	Equation 1-5	%%%%%%%%%%%%%%%%%%%%%%%%%%%%%%%%%
\begin{eqnarray}
V_{1+}^{\; \text{eff}} &=& \frac{1}{\Delta} \left(\varepsilon_{\text{et}} V_{1+}^{\; \text{ext}} + v \Pi^0_{\,2}(1-2G_{12})V_{2+}^{\; \text{ext}}  \right)  \label{eq1}\\
V_{2+}^{\; \text{eff}} &=& \frac{1}{\Delta} \left(\varepsilon_{\text{ht}} V_{2+}^{\; \text{ext}} + v \Pi^0_{\,1}(1-2G_{12})V_{1+}^{\; \text{ext}}  \right)  \label{eq2}\\
\Delta &=& \varepsilon_{\text{et}} \varepsilon_{\text{ht}} - v^2 \Pi^0_{\,1} \Pi^0_{\,2} (1-2G_{12})^2 \label{eq3}
\end{eqnarray}
\begin{eqnarray}
%\Delta &=& \varepsilon_{\text{et}} \varepsilon_{\text{ht}} - v^2 \Pi^0_{\,1} \Pi^0_{\,2} (1-2G_{12})^2 \\ 
V_{1-}^{\; \text{eff}} &=& \frac{V_{1-}^{\; \text{ext}} }{(1-G_{1-}v\Pi^0_{\,1})}  \label{eq4}  \\
V_{2-}^{\; \text{eff}} &=& \frac{V_{2-}^{\; \text{ext}} }{(1-G_{2-}v\Pi^0_{\,2})}   \label{eq5}
\end{eqnarray}
%%%%%%%%%%%%%%%%%%%%%%%%%%%%%%%%%%%%%%%%%%%%%%%%%%%%%%%%%%%%%%%%%%%%%
These equations are written in a suggestive form to show the close relation to the separate electron and fermion gases. 

The mean field linear density response is given by $\Delta n_{2+} = -\Pi^{0}_2 V_{2+}^{\; \text{eff}}$ where $\Pi^{0}_2$ is the electron Lindhard function and the electron test charge dielectric function is $\varepsilon_{\text{et}} = 1 + v (1-G_{2+}) \Pi^{0}_2$, where $G_{2+}$ is the electron-electron local field factor which includes exchange and correlation. Similarly, $\Delta n_{1+} = -\Pi^{0}_1 V_{1+}^{\; \text{eff}}$, and $\varepsilon_{\text{ht}} = 1 + v (1-G_{1+}) \Pi^{0}_{\,1}$.

These equations are quite general, but the local field factors are generally unknown. They are identical to the results in Refs \cite{ref3,ref4} with the exception of the factor of two in the definition of the electron-positive fermion local field factor $G_{12}$.

The key to using these equations is knowledge of the local field factors, and a careful definition of the external potentials. In the general case, an electron or positive fermion with spin up or spin down can have different interactions with an external potential. For a simple coulomb test charge, both spins will have the same interaction and the total induced density will be the sum of the spin-up and spin-down electron densities. That is denoted by the $+$ subscript. The minus $-$ subscript refers to the difference in the induced densities of spin-up and spin-down electrons. For a coulomb external potential, the number of induced electrons with spin-up is equal to the number with spin-down, and therefore the quantities with a minus-subscript are zero. A purely coulomb disturbance does not induce a magnetic response.

Similarly, a magnetic field will have the opposite effect on opposite spins. The sum of the induced densities will be zero, but the difference will be finite. A magnetic disturbance does not induce a density response. Equations \eqref{eq4} \& \eqref{eq5} show that the magnetic response of the paramagnetic system is simply the sum of the magnetic responses of the electrons and positive fermion separately. If there were two pockets of electrons that could exchange with each other, this could lead to magnetic instabilities such as ferromagnetism and spin density waves. This is not considered in this paper.

However, if the disturbance is an electron or positive fermion with spin-up, it will induce both a density and magnetic response. This is discussed in the following paper.

It is not easy to get an intuitive feeling from Eqs. (\ref{eq1}-\ref{eq5}). For a perturbation that is a positive test charge $v$, $V_{1+}^{\;\text{ext}} = v = - V_{2+}^{\text{ext}}$, and Eq. \eqref{B15} can be rewritten as 
%%%%%%%%%%%%%%%%%%%%	Equation 6	%%%%%%%%%%%%%%%%%%%%%%%%%%%%%%%%%
\begin{eqnarray}\label{eq6}
V_{2+}^{\text{eff}} = \frac{\frac{-v}{\varepsilon_{\text{et}}} \left( 1+ v \Pi^0_{\,1}\frac{(1-2G_{12})}{\varepsilon_{\text{ht}}}\right)}{1-v^2 \Pi^0_{\,1} \Pi^0_{\,2} \frac{(1-2G_{12})^2}{\varepsilon_{\text{et}} \varepsilon_{\text{ht}}}}  \; .
\end{eqnarray}
%%%%%%%%%%%%%%%%%%%%%%%%%%%%%%%%%%%%%%%%%%%%%%%%%%%%%%%%%%%%%%%%%%%%%
If there were no positive fermions, the effective interaction would just be the external potential divided by the electron test charge dielectric function: $v / \varepsilon_{\text{et}}$. The second term in the numerator is the additional induced charge due to the positive fermion response that reduces the effective charge that the electron sees. The additional term in the denominator reflects the modifications to $\varepsilon_{\text{et}}$ due to the positive fermions. There is additional screening by the positive fermions.

It is well known in the uniform electron gas that $v / \varepsilon_{\text{et}}$ is completely well-behaved and has no singularities that would imply a phase transition.                                                                    

The electron test charge dielectric function $\varepsilon_{\text{et}} = 1 + v (1-G_{2+}) \Pi^0_2$ was earlier speculated to have a potential charge density wave when $\varepsilon_{\text{et}} =0$ if $G_{2+}$ became significantly larger than $1$ at $q < 2k_{F}$ before the Lindhard function started to decrease rapidly (see the discussion on the “hump” in Ref. \cite{ref11}). Sum rules and Quantum Monte Carlo calculations have accurately determined $G_{2+}(q)$ and there are no charge density waves in the electron gas in a uniform positive background. Similarly, there are no spin density waves \cite{ref14}.

The effect of the additional screening by the positive fermions is essentially to modify the electron-test charge dielectric function to be of the form $1 + v (A - G_{2+})\Pi^0_2$   with $A=1$ for the uniform electron gas and $A<1$ in the electron-positive fermion gas. For small enough $A$, this term can and does become zero at finite $q$. This is the origin of the charge density waves.

When the external potential is a positive coulomb potential, Eq. \eqref{eq2} and Eq. \eqref{eq3} become
%%%%%%%%%%%%%%%%%%%% Equation 7 - 8 %%%%%%%%%%%%%%%%%%%%%%%%%%%%%%%%%%
\begin{eqnarray}
    V_{2+}^{\text{eff}} &=& -\frac{v}{\Delta} \left( \varepsilon_{\text{ht}} - v\Pi^0_1(1-2G_{12}) \right) \nonumber\\
    &=& -\frac{4\pi e^2}{q^2 \Delta} \left( 1 - v\Pi^0_1(G_{1+} - 2G_{12}) \right) \; , \label{eq7} \\
    V_{1+}^{\text{eff}} &=& \frac{v}{\Delta} \left( \varepsilon_{\text{et}} - v\Pi^0_2(1-2G_{12}) \right) \nonumber \\
    & = & \frac{4\pi e^2}{q^2 \Delta} \left( 1 - v\Pi^0_2(G_{2+} - 2G_{12}) \right) \; . \label{eq8}
\end{eqnarray}
%%%%%%%%%%%%%%%%%%%%%%%%%%%%%%%%%%%%%%%%%%%%%%%%%%%%%%%%%%%%%%%%%%%%%%
The instabilities occur when these effective potentials and therefore the induced densities diverge. The instabilities are all contained in $\Delta$ and the rest of this section focuses on these instabilities.

The expression for $\Delta$ can be rewritten to show that terms in $\varepsilon_{et}$ $\varepsilon_{ht}$ exactly cancel the last term which varies as $1/q^4$ at small $q$,
%%%%%%%%%%%%%%%%%%%% Equation 9 - 12 %%%%%%%%%%%%%%%%%%%%%%%%%%%%%%%%%%
\begin{eqnarray}
    \Delta &=& \varepsilon_{et}  \varepsilon_{ht} - v^2 \Pi^0_1 \Pi^0_2 (1-2 G_{12})^2 \; ,\\
    \nonumber\\
    \varepsilon_{et} &=& 1 + v(1-G_{2+})\Pi^0_2 \nonumber \\
    &=& 1 - v \Pi^0_2(G_{2+} \! \!- 2G_{12}) + v (1 - 2 G_{12})\Pi^0_2 \; , \\
    \nonumber\\
    \varepsilon_{ht} &=& 1 + v(1-G_{1+})\Pi^0_1 \nonumber \\
    &=& 1 - v \Pi^0_1(G_{1+} \! \!- 2G_{12}) + v (1 - 2 G_{12})\Pi^0_1 \; , \\
    \nonumber\\
    \Delta &=& \left( 1 \! - v(G_{2+}\!\! - 2G_{12})\Pi^0_2 \right) \left( 1 \! - v(G_{1+} \!\!- G_{12})\Pi^0_1 \right)\nonumber \\
    && \; \; + \left( 1 - v (G_{2+} -2 G_{12})\Pi^0_2 \right)v\Pi^0_1\\
    && \; \; \; + \left( 1 - v (G_{1+} -2 G_{12})\Pi^0_1 \right)v\Pi^0_2 \label{eq12}\nonumber
\end{eqnarray}
%%%%%%%%%%%%%%%%%%%%%%%%%%%%%%%%%%%%%%%%%%%%%%%%%%%%%%%%%%%%%%%%%%%%%
This is a general result from linear response theory. In the simple model of this paper the electron-positive fermion correlation energy is neglected and this implies that $G_{12} =0$, and $(1 – v G_{2+}\Pi^0_2) = \kappa_{02}/\kappa_2$ at $q=0$ through the compressibility sum rule. $\kappa = 1/B$ is the compressibility. $\kappa_0$ is the compressibility of non-interacting fermions. I expect that a more complete theory has a similar sum rule.

At small $q$, the first term in Eq. \eqref{eq12} is a constant and the next two terms are proportional to $1/q^2$. The denominator in Eqs. (\ref{eq7} \& \ref{eq8}) at $q=0$ is 
%%%%%%%%%%%%%%%%%%%% Equation 13 %%%%%%%%%%%%%%%%%%%%%%%%%%%%%%%%%%
\begin{eqnarray}\label{eq13}
    q^2 \Delta &=& q_{T\!F2}^2 \frac{\kappa_{01}}{\kappa_1} + q_{T\!F1}^2 \frac{\kappa_{02}}{\kappa_2}  \\
    &=& q_{T\!F2}^2 \kappa_{02} \frac{M}{m} \left( \frac{1}{\kappa_1} + \frac{1}{\kappa_2} \right) = q_{T\!F2}^2 \kappa_{02} \frac{M}{m}B \nonumber
\end{eqnarray}
%%%%%%%%%%%%%%%%%%%%%%%%%%%%%%%%%%%%%%%%%%%%%%%%%%%%%%%%%%%%%%%%%%%%
where $B = B_1 + B_2$ is the total bulk modulus (inverse of the compressibility) of the electron-positive fermion gas. The scalings $q_{TF1}^2 = (M/m) q_{TF2}^2$ and $B_{01} = m/M B_{02}$ were used where $B_{02}= 2/3 n \varepsilon_{F2}$ and $n$ is the density and $\varepsilon
_{F2}$ is the Fermi energy of the electrons which is proportional to $1/m$.

Equation \eqref{eq13} shows a universal divergence at $B=0$. This combined with Fig. \ref{fig1} shows that the electron-positive fermion gas has a true instability at a certain $r_s$ for all mass ratios $M/m$. This is called the compressibility instability. Near the instability, $B_2$ the bulk modulus of the electrons is positive and $B_1$ the bulk modulus of the positive fermions is negative. Therefore it is the electrons that are providing stability to the system.

The effective potentials and induced densities will be discussed after investigation of the finite $q$ the divergence of $q^2\Delta$, which signifies the charge density waves, and the resulting overall phase diagram of the electron-positive fermion gas.

When the electron-positive fermion correlation energy becomes available, the new bulk modulus will be used in the compressibility sum rule, and I expect $G_+$ will be replaced by $G_+ - 2G_{12}$ in the response functions. This will determine $G_{12}$ at $q=0$, but not at finite $q$.

To understand the charge density divergence, I rewrite the expression for $\Delta$ with $G_{12}=0$ as
%%%%%%%%%%%%%%%%%%%% Equation 14 %%%%%%%%%%%%%%%%%%%%%%%%%%%%%%%%%%
\begin{eqnarray}\label{eq14}
    \Delta &=& \varepsilon_{\text{ht}} \left( \varepsilon_{\text{et}} - v^2 \frac{\Pi^0_1 \Pi^0_2}{\varepsilon_{\text{ht}}}  \right) \\
    &=& \varepsilon_{\text{ht}} \left( 1 + \left( 1 - G_{2+}-v \frac{\Pi^0_1}{\varepsilon_{\text{ht}}}   \right) v \Pi^0_2         \right) \; . \nonumber
\end{eqnarray}
%%%%%%%%%%%%%%%%%%%%%%%%%%%%%%%%%%%%%%%%%%%%%%%%%%%%%%%%%%%%%%%%%%%%
This will diverge if $(1+ (1- G_{2+} -v \Pi^0_1/\varepsilon_{\text{ht}} ) v \Pi^0_2) = 0$. It is known from the electron gas that $\varepsilon_{\text{et}} = 1+(1 - G_{2+} v \Pi^0_2) > 0$ and the same is true for $\varepsilon_{\text{ht}}$ and neither diverge. Therefore it is the additional screening $-v \Pi^0_1/\varepsilon_{\text{ht}}$ from the positive fermions that drives the divergence and charge density waves.

The discussion below will show that for certain mass ratios $M/m$, there is a divergence at intermediate $q/k_F$ where the Lindhard function is close to one and slowly varying. There is also another divergence for large mass ratios as $q/k_F$ approaches $2$, where the Lindhard function begins to fall rapidly through its logarithmic singularity which is magnified by the factor of $M/m$ and is responsible for this divergence. The local field factor begins changing rapidly near $q/2k_F$ and details of the transition here are probably not reliable.                                       

In order to make quantitative predictions, one needs to know the local field factors. The approximation I use is to assume that the electron local field factors are the same as those in the uniform electron gas, and the positive fermion local field factors and Lindhard function are scaled from the electron gas to the mass of the positive fermion. The scaling is derived in Refs. \cite{ref12,ref15}  The electron - positive fermion local field factor $G_{12}$ is assumed to be zero. Furthermore, I use the simplified electron gas local field factors which are quite accurate up to $q/k_F=2$. Use of local field factors with greater precision such as those in Ref. \cite{ref17} will make only a small quantitative difference. The exact expressions for the local field factors used in the calculations are given in Appendix C. The effective $r_s$ of the positive fermions is $M/m$ times that of the electrons.

The electron-positive fermion system is unstable when $\Delta = 0$. It is convenient to use $\Delta \cdot(q/k_F)^2$ which is in the denominator of the effective potentials. With the assumptions of the simple model, $\Delta \cdot(q/k_F)^2$ can be evaluated at all densities, mass ratios $M/m$ and wave vectors. In addition, the net induced density can be calculated in the standard linear response approach. 

An important point is that the instabilities depend strongly on the local field factor $G_{2+}$. The small q behavior is determined by the compressibility sum rule, and all versions of the local field factor will yield the same $q=0$ instability if they satisfy the compressibility sum rule. The finite $q$ instabilities depend strongly on the $q$ dependence of the local field factor.

In general, the wave vector dependence of the local field factor is difficult to calculate, and also difficult to physically understand. The compressibility sum rule determines the behavior at small $q$ as
%%%%%%%%%%%%%%%%%%%%%%% Ecuacion 15 #%%%%%%%%%%%%%%%%%%%%%%%%%%%%%%%
\begin{eqnarray}\label{eq15}
    G_{2+}(q) &=& \left(  1- \frac{\kappa_{02}}{\kappa_2}  \right)\left( \frac{q}{q_{T \! F2}}\right)^2 \; .
\end{eqnarray}
%%%%%%%%%%%%%%%%%%%%%%%%%%%%%%%%%%%%%%%%%%%%%%%%%%%%%%%%%%%%%%%%%%%
Early theories predicted that the local field factor should become a constant at large $q$ and Hubbard suggested a form for the $q$ dependence, which when corrected for the compressibility sum rule is 
%%%%%%%%%%%%%%%%%%%%%%%%%%%%    Ecuacion16    %%%%%%%%%%%%%%%%%%%%%%%
\begin{eqnarray}\label{eq16}
    G_{\text{Hubb}}(q) &=& \displaystyle \frac{q^2}{\displaystyle 2q^2 + \frac{\qtf^2}{\displaystyle 1-\frac{\kappa_{02}}{\kappa_2}}}
\end{eqnarray}
%%%%%%%%%%%%%%%%%%%%%%%%%%%%%%%%%%%%%%%%%%%%%%%%%%%%%%%%%%%%%%%%%%%%
which satisfies the compressibility sum rule and goes to the constant $1/2$ at large $q$. Other versions of the Hubbard approximation approached different constants at large $q$.

Later theories predicted that the local field factor also behaves as $q^2$ at large $q > 2k_F$, but with a different coefficient. The behavior of the local field factor at intermediate $q$ near $2k_F$ was the subject of considerable research. Quantum Monte Carlo calculations confirmed the compressibility sum rule and established that the local field factor continued to increase approximately as $q^2$ until about $2k_F$. The Quantum Monte Carlo calculations and the predicted large $q$ behavior were connected by interpolation, and the resulting local field factors are given in Ref. \cite{ref14}. The Lindhard function is equal to one at small $q$, and falls off quickly near $q=2k_F$. The falloff of the Lindhard function justifies the use of a simpler version of the local field factor in Eq. \eqref{eq15}. Use of the better local field factor from Ref. \cite{ref14} will make a modest quantitative difference.

\begin{figure}[h!]
	\centering
    \includegraphics[width=1.0\columnwidth]{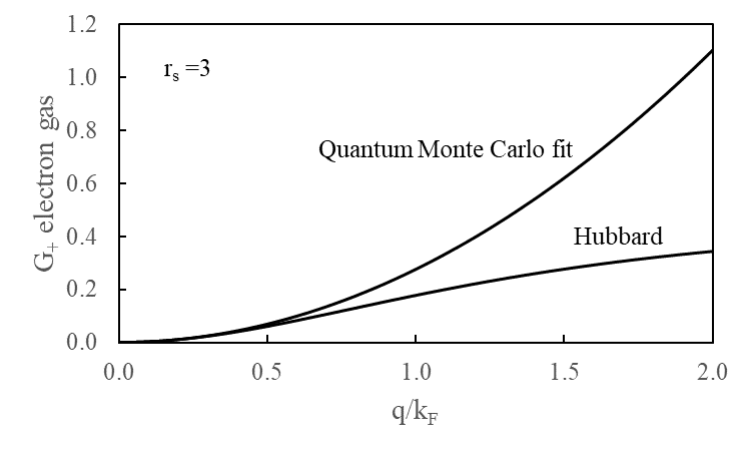}
    \caption{The accurate local field factor $G_+$ in the uniform electron gas at $r_s = 3$ determined by Quantum Monte Carlo calculations compared to the earlier Hubbard approximation.}
    \label{fig2}
	\end{figure}

The local field factor in Eq. \eqref{eq15} that satisfies the compressibility sum rule and fits the Quantum Monte Carlo data is compared to the Hubbard approximation Eq. \eqref{eq16} in Fig. \ref{fig2}.

%%%%%%%%%%55%%%%%%55	Fig 2	%%%%%%%%%%%%%%%%%%%%%%%%%%%%%%%%%%%%%
%%%%%%%%%%%%%%%%%%%%%%%%%%%%%%%%%%%%%%%%%%%%%%%%%%%%%%%%%%%%%%%%%%%%%

No finite $q$ (charge density wave) instabilities occur using the Hubbard approximation, because the value of the Hubbard approximation near $q \approx k_F$ is too small. This shows the importance of the intermediate $q$ behavior of the local field factor. Using the accurate local field factor Eq. \eqref{eq15} predicts charge density waves for mass ratios $M/m > 4.97$  at wave vectors above $q/k_F=0.83$. This demonstrates the importance of the local field factor at intermediate wave vectors.

%I have calculated the denominator of the induced density $\Delta * (q/k_F)^2$  and the induced density $\Delta n_1 + \Delta n_2 $. For simplicity in the notation, I have dropped the subscript $+$ from the induced densities. 

The behavior of $\Delta * (q/k_F)^2$ is somewhat complicated and I examine it in detail for several mass ratios $M/m = 1.2, \, 3, \, 4.97, \, 6, \, 10$. At each mass ratio, the density $r_s$ is varied to find the values where $\Delta * (q/k_F)^2 = 0$. 

%%%%%%%%%%55%%%%%%55	Fig 3	%%%%%%%%%%%%%%%%%%%%%%%%%%%%%%%%%%%%%
\begin{figure}[h!]
	\centering
    \includegraphics[width=1.0\columnwidth]{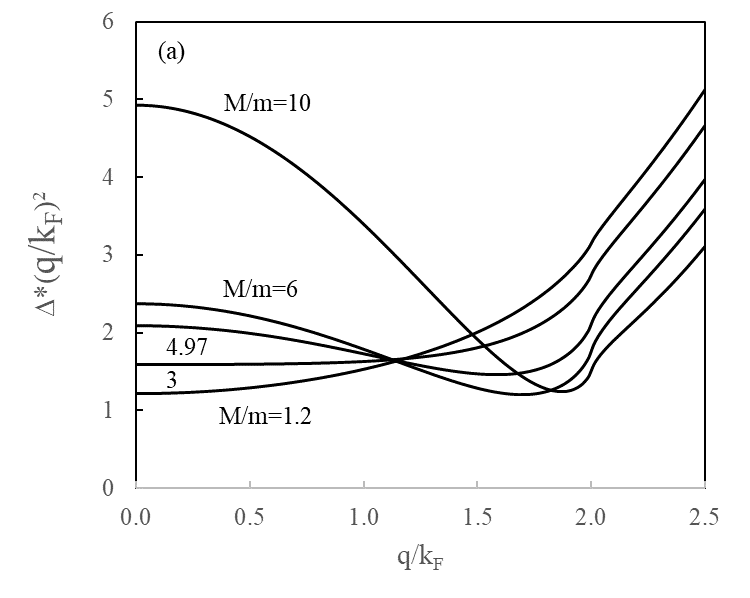}
    \includegraphics[width=1.0\columnwidth]{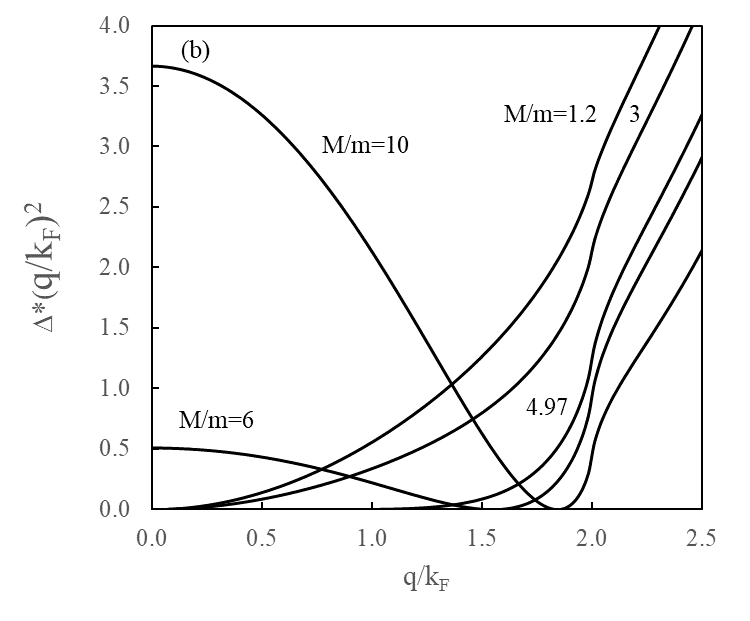}
    \caption{Plot of $\Delta * (q/k_F)^2$, the denominator of the induced density as a function of $q/k_F$ for representative values of the mass ratio $M/m$. (a) shows the value at the energy minimum of the electron-positive fermion gas; (b) shows points where the denominator become zero and the system becomes unstable.}
    \label{fig3}
	\end{figure} 
%%%%%%%%%%%%%%%%%%%%%%%%%%%%%%%%%%%%%%%%%%%%%%%%%%%%%%%%%%%%%%%%%%%%%

Figure \ref{fig3}(a) shows that at the $r_s$ of the energy minimum, the denominator is well above zero and reasonably well behaved. There is no instability. However if there is a pseudo potential or other mechanism to reduce the density to approximately $1.25$ times the $r_s$ of the energy minimum, the denominator become zero, and system is unstable at $q=0$. For $1 < M/m < 4.97$, the instability occurs at $q=0$. %The electron-positive fermion gas may condense in a portion of the volume, or other transitions such as droplets or excitons. 
At $M/m=4.97$, the denominator has simultaneous minima at $q/k_F=0$ and $q/k_F=0.83$ which signifies a charge density wave. This is a triple point. For $M/m > 4.97$, the denominator first becomes zero at a finite value of $q/k_F$ and the system has a charge density wave. The charge density wave appears at smaller $r_s$ than that of the $q=0$ instability. Fig \ref{fig3}(b) shows the behavior near the critical density at each mass ratio.

%%%%%%%%%%55%%%%%%55	Fig 4	%%%%%%%%%%%%%%%%%%%%%%%%%%%%%%%%%%%%%
\begin{figure}[h!]
	\centering
    \includegraphics[width=1.0\columnwidth]{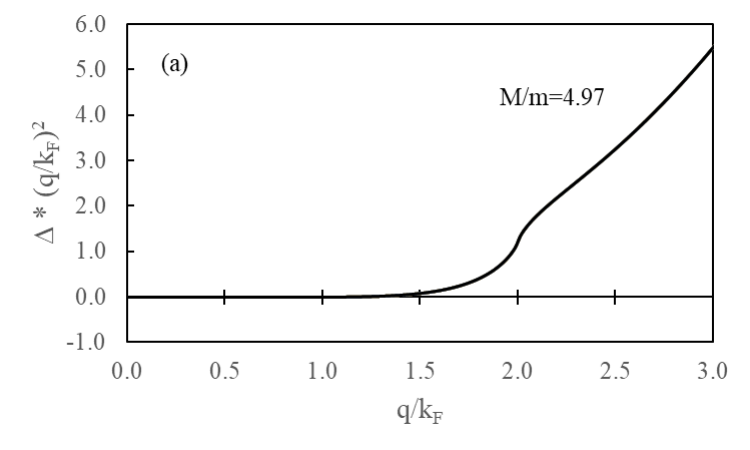}
    \includegraphics[width=1.0\columnwidth]{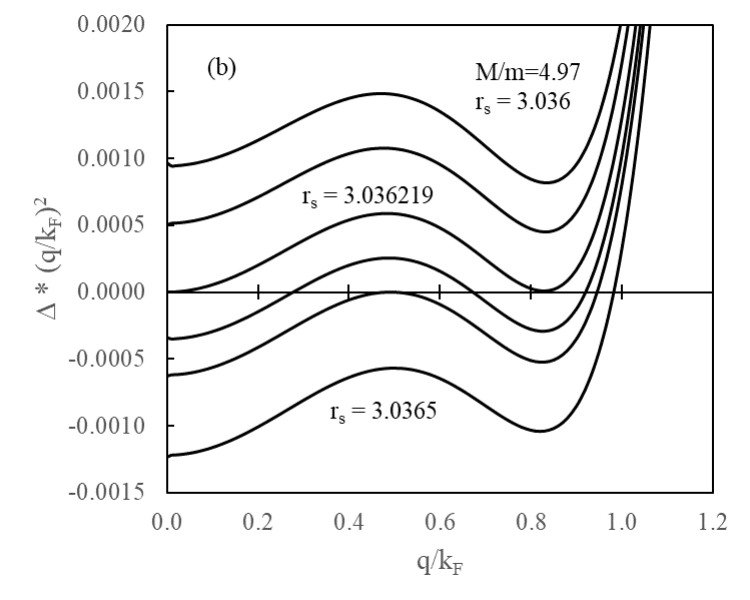}
    \caption{Plot of $\Delta * (q/k_F)^2$, the denominator of the induced density as a function of $q/k_F$ for $M/m=4.97$ at densities near the triple point at $r_s = 3.036219$. $\Delta * (q/k_F)^2$ at a gross scale is shown in (a), and at 100 times magnification in (b).}
    \label{fig4}
	\end{figure} 
%%%%%%%%%%%%%%%%%%%%%%%%%%%%%%%%%%%%%%%%%%%%%%%%%%%%%%%%%%%%%%%%%%%%%

Figure \ref{fig4}(a) shows that the denominator $\Delta * (q/k_F)^2$ for mass ratio $M/m =4.97$ on a gross scale is quite flat and near zero from   $0 < q/k_F <1$. The magnified view in (b) details the complicated behavior as a function of density $r_s$. The triple point is the point where the a stable electron-positive fermion gas, a condensation instability at $q=0$, and a charge density wave at finite $q$ coexist. In the region of the triple point, there is interesting behavior in a very small region of density. At higher density (smaller $r_s$) $\Delta * (q/k_F)^2$ is positive and the gas is stable. At the triple point, $\Delta * (q/k_F)^2$ is simultaneously zero at $q/k_F = 0$ and $0.83$. At very slightly lower densities (larger $r_s$), $\Delta * (q/k_F)^2$ is negative near $q/k_F = 0$, but becomes positive for a small region before becoming negative again and then positive near $q/k_F=1$. At very slightly lower density, this small positive region disappears. 

The extreme sensitivity to density shown in Fig. \ref{fig4}b is apparent because the entire range of $r_s$ is $\pm 0.01 \%$ around the ``magic" value at the triple point. Recall however that at $M/m = 4.97$, the $r_s$ at the energy equilibrium is $2.43$ and at that density the system is stable and featureless as shown in Fig. \ref{fig3}a.

%%%%%%%%%%55%%%%%%55	Fig 5	%%%%%%%%%%%%%%%%%%%%%%%%%%%%%%%%%%%%%
\begin{figure}[h!]
	\centering
    \includegraphics[width=1.0\columnwidth]{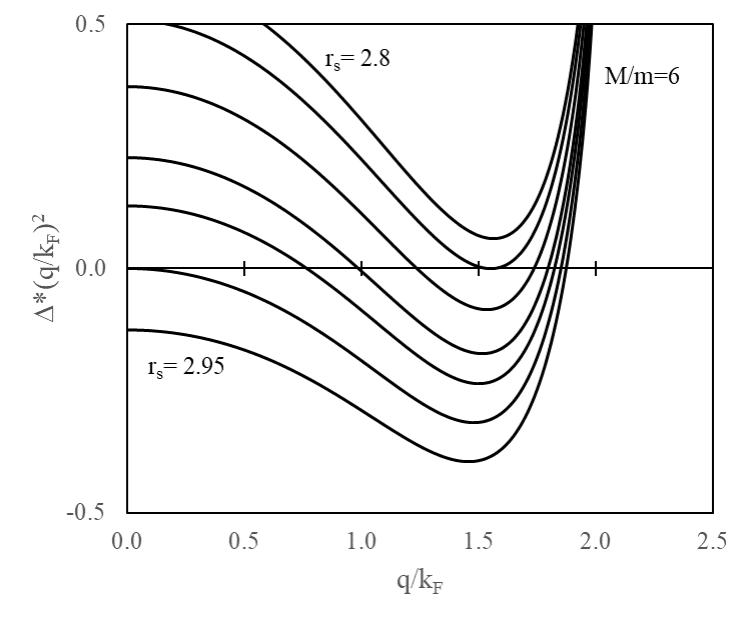}
    \caption{Plot of $\Delta * (q/k_F)^2$, the denominator of the induced density as a function of $q/k_F$ at $M/m=6$ near the charge density wave at $r_s = 2.821$ and the $q = 0$ instability at $r_s = 2.9256$.}
    \label{fig5}
	\end{figure} 
%%%%%%%%%%%%%%%%%%%%%%%%%%%%%%%%%%%%%%%%%%%%%%%%%%%%%%%%%%%%%%%%%%%%%

Figure \ref{fig5} shows that $\Delta * (q/k_F)^2$ at the mass ratio $M / m = 6$ is always positive and the electron-positive fermion gas is completely stable for $r_s < 2.821$ when the charge density wave appears. In the region $2.821<  r_s < 2.9256$, $\Delta * (q/k_F)^2$ is positive for wave vectors below the critical wave vector of the charge density wave. For lower densities (larger $r_s$), $\Delta * (q/k_F)^2$ does not become positive until near $q/k_F=1.5$. 

%%%%%%%%%%55%%%%%%55	Fig 6	%%%%%%%%%%%%%%%%%%%%%%%%%%%%%%%%%%%%%
\begin{figure}[h!]
	\centering
    \includegraphics[width=1.0\columnwidth]{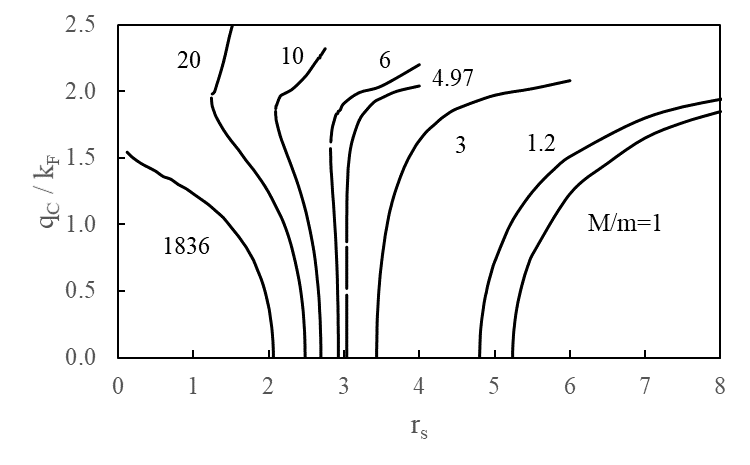}
    \caption{The critical wave vector, $q_C/k_F$ , is defined as the locus of points where $\Delta * (q/k_F)^2 =0$ as a function of $r_s$ for a given mass ratio. Note that there can be two values of $q_C / k_F$ at a given $r_s$ (or three at the triple point) where $\Delta * (q/k_F)^2$ crosses zero two or three times. The small portions of missing data reflect where the two values merge and it is more difficult to calculate. Small choppiness in the curves is due to limited accuracy in the calculation.}
    \label{fig6}
	\end{figure} 
%%%%%%%%%%%%%%%%%%%%%%%%%%%%%%%%%%%%%%%%%%%%%%%%%%%%%%%%%%%%%%%%%%%%%

Figure \ref{fig6} represents the phase diagram of the electron-positive fermion gas. The gas with a mass ratio of $M/m$ and density $r_s$ is stable against a disturbance of wave vector $q$ if $q < q_C$. In the simple model, $\Delta * (q/k_F)^2$ is completely specified for all values of wave vector, density and mass ratio. The three-dimensional surface is complicated and views from several different perspectives are useful.

%%%%%%%%%%55%%%%%%55	Fig 7	%%%%%%%%%%%%%%%%%%%%%%%%%%%%%%%%%%%%%
\begin{figure}[h!]
	\centering
    \includegraphics[width=1.0\columnwidth]{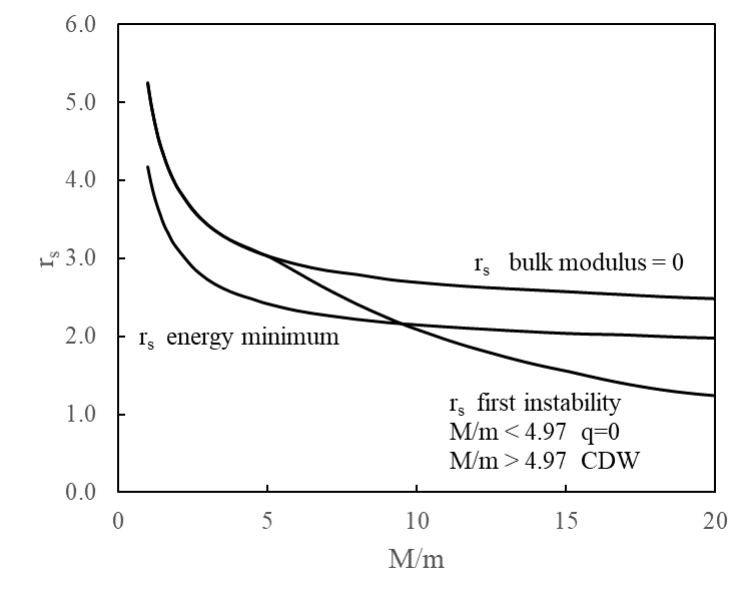}
    \caption{Critical density $r_s$ of the first instability encountered compared to the critical density where the bulk modulus equals zero in the simple model. For comparison the $r_s$ of the energy minimum in the simple model is also shown.}
    \label{fig7}
	\end{figure} 
%%%%%%%%%%%%%%%%%%%%%%%%%%%%%%%%%%%%%%%%%%%%%%%%%%%%%%%%%%%%%%%%%%%%%

For each mass ratio $M/m$, the electron-positive fermion gas has a unique energy minimum and $q=0$ instability point which is at closely $1.25$ times the $r_s$ of the energy minimum as shown in Fig \ref{fig1}. The other curve plotted in Fig. \ref{fig7} is the value of $r_s$ that is the minimum value where $\Delta * (q/k_F)^2 = 0 $ for a given $M/m$. This can be understood from Fig. \ref{fig6}. This minimum $r_s$ is at different critical wave vectors for different mass ratios. For example at $M/m=1.2$, the minimum value of $r_s= 4.7997$ at $q/k_F=0$. Likewise at $M/m=10$ the minimum value of $r_s= 2.09192$ at $q/k_F=1.85$. This is the density where a charge density wave first appears for the uniform gas.

%%%%%%%%%%%%%%%%%%%%%%%%%%%%%%%%%%%%%%%%%%%%%%%%%%%%%%%%%%%%%%%%%%%%%%%%%%%%%%%%%%%%%%%%%%%%%%%%%%%%%%
%Note that the $r_s$ of the energy minimum (equilibrium position) is below the minimum $r_s$ for a charge density wave until approximately $M/m=9$ indicating a totally stable gas. Above that mass ratio, the energy minimum $r_s$ is below the minimum for the $q=0$ instability, and above that for a charge density wave, indicating a region of wave vector where $\Delta * (q/k_F)^2$ is positive before reaching the critical wave vector of the charge density wave.
%%%%%%%%%%%%%%%%%%%%%%%%%%%%%%%%%%%%%%%%%%%%%%%%%%%%%%%%%%%%%%%%%%%%%%%%%%%%%%%%%%%%%%%%%%%%%%%%%%%%%%%
From  Figs. \ref{fig6} \& \ref{fig7}, it can be seen that for each mass ratio equal to or above $M/m=4.97$, a charge density wave is predicted at a certain $r_s$ with a specific wave vector $q_c$ and that the charge density wave occurs at a higher density (smaller $r_s$) then the compressibility instability at $q=0$.

For $4.97<M/m<9$, the $r_s$ of the energy minimum is below that of both the $q=0$ and charge density wave instabilities and the system is predicted to be stable and without a charge density wave. For $M/m>9$, the system at the $r_s$ of the energy minimum is at a lower density than that where a charge density wave is predicted. The electron-positive fermion gas at the energy minimum would have a charge density wave for all mass ratios $M/m>9$. 
%Of course the charge density wave would contradict the basic assumption of a uniform background. I do not understand what this means. If the amplitude of the charge density wave is small, perhaps the energy would not be significantly affected and the equilibrium $r_s$ would not be significantly changed, and the system would have an intrinsic charge density wave. This point merits investigation.
This confusing situation is a clear prediction of the simple model and results from the assumption that the local field factors for the electrons and positive fermions grow as $q^2$ up to $q = 2k_F$. The validity of this assumption is discussed in Section V. Of course, the charge density wave invalidates the basic assumption of a uniform background. If the amplitude of the charge density wave is small, perhaps the energy would not be significantly affected, and the equilibrium $r_s$ would not be significantly changed and the system would have an intrinsic charge density wave. This prediction definitely merits further investigation. The remainder of the paper continues to examine the predictions of the simple model using the assumed local field factors.

I have calculated the instability $\Delta * (q/k_F)^2 = 0 $ for a range of $M/m$ from $1-290$. In all cases the system becomes unstable at a certain $r_s$. For $1 <M/m< 4.97$, the instability is at $q=0$. For $4.97 < M/m < 255$, the instability is at finite critical wave vector which ranges from $0.82 < q/k_F < 1.99$ as $M/m$ increases. This instability is a charge density wave resulting from the additional screening from the positive fermions. This charge density wave instability occurs because the local field factor continues to increase approximately as $q^2$ up to $2k_F$. At $M/m = 255$, there appears another abrupt transition which is due to the logarithmic singularity in the Lindhard function which is magnified by the mass ratio $M/m$ in $\varepsilon_{\text{ht}}$. For larger $M/m$, the critical wave vector appears at $q/k_F = 2.18$, and the instability is again a charge density wave. As stated earlier, the local field factor and the Lindhard function are both changing quickly with wave vector and the results near $q/k_F=2$ should be viewed as qualitative. There is certainly another transition, but the predicted mass ratio and density are qualitative because the local field factor has the most uncertainty in this region.

%%%%%%%%%%55%%%%%%55	Fig 8	%%%%%%%%%%%%%%%%%%%%%%%%%%%%%%%%%%%%%
\begin{figure}[h!]
	\centering
    \includegraphics[width=1.0\columnwidth]{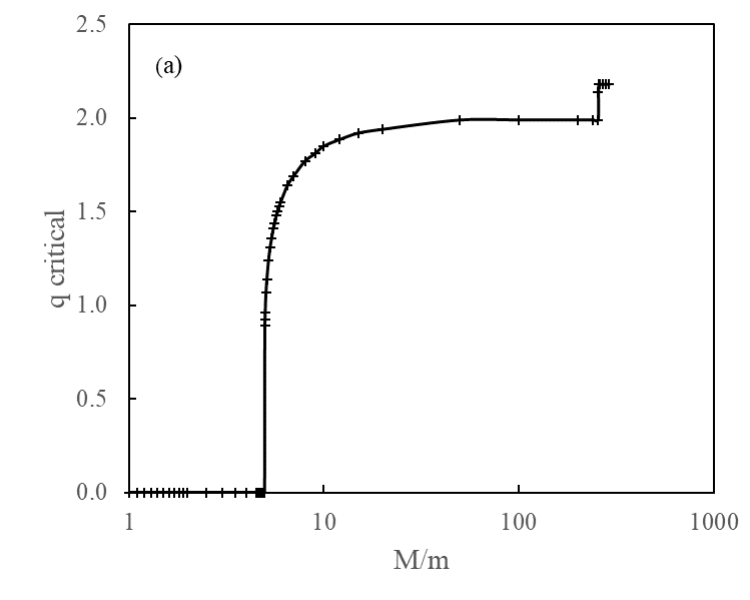}
    \includegraphics[width=1.0\columnwidth]{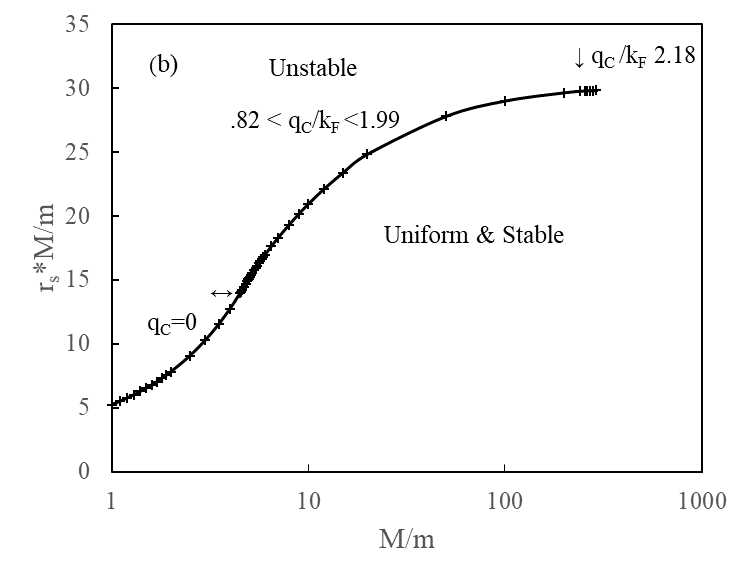}
    \caption{(a) First unstable wave vector, when decreasing density (increasing $r_s$) and (b) $r_s$ at that wave vector of the electron-positive fermion gas as a function of the ratio of the positive fermion mass to the electron mass.}
    \label{fig8}
	\end{figure} 
%%%%%%%%%%%%%%%%%%%%%%%%%%%%%%%%%%%%%%%%%%%%%%%%%%%%%%%%%%%%%%%%%%%%%

%%%%%%%%%%55%%%%%%55	Fig 9	%%%%%%%%%%%%%%%%%%%%%%%%%%%%%%%%%%%%%
\begin{figure}[h!]
	\centering
    \includegraphics[width=0.95\columnwidth]{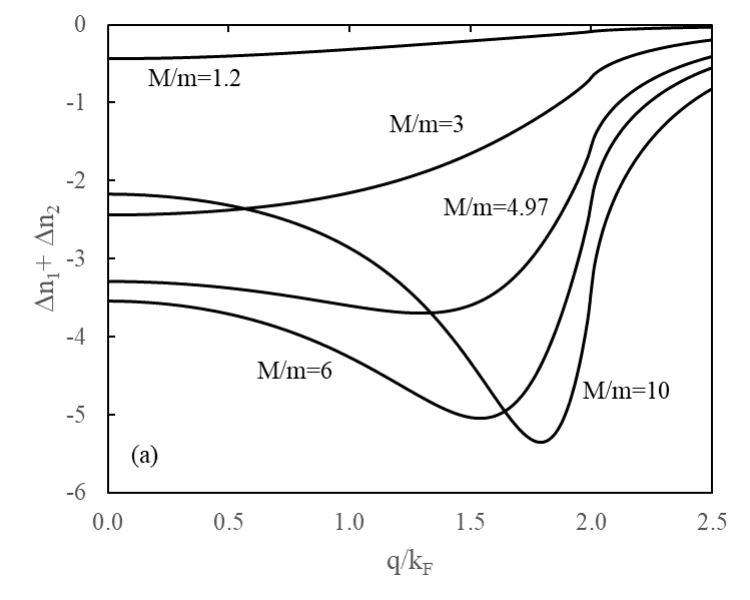}
    \includegraphics[width=0.95\columnwidth]{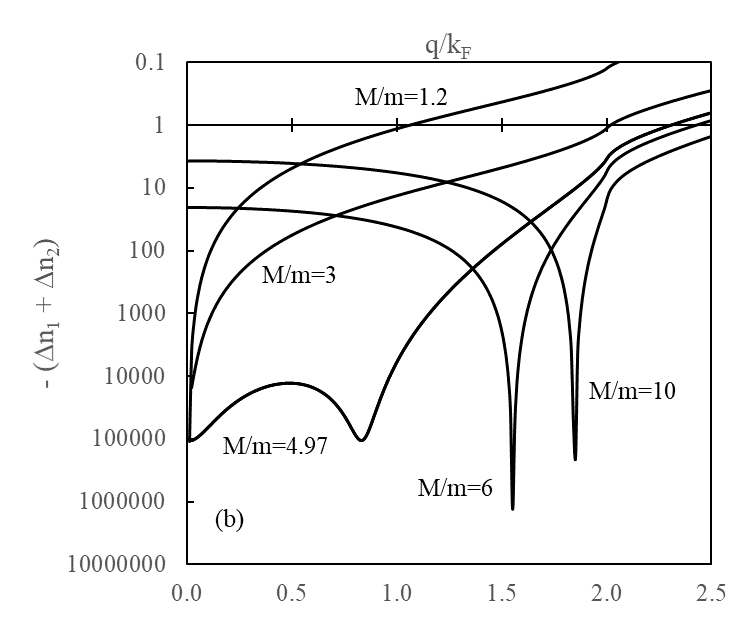}
    \caption{The total induced density $\Delta n_1 + \Delta n_2 $ due to a positive test charge disturbance for representative mass ratios $M/m$. (a) is at the energy minimum and has a linear scale and (b) is just below the point of instability which is $q=0$ for $M/m < 4.97$ and at finite $q$ for $M/m > 4.97$ and at both for $M/m=4.97$ with a logarithmic scale.}
    \label{fig9}
	\end{figure} 
%%%%%%%%%%%%%%%%%%%%%%%%%%%%%%%%%%%%%%%%%%%%%%%%%%%%%%%%%%%%%%%%%%%%%

Figure \ref{fig8} is virtually identical to Fig. 3 of Ref. \cite{ref12} where the authors used density functional theory within the framework of the local spin density approximation.  They also neglected the additional correlation effects between the electrons and positive fermions. The close agreement is a validation of the simple model that yields the same results using Microsoft Excel on a laptop. Ref. \cite{ref12} has a good discussion and suggests connections to possible exotic phase transitions, new intermediate phases and smectic liquid crystals. I do not have an understanding of what is occurring on the other side of instabilities. The agreement between the simple model and density function theory should have been expected, because the density functional theory used the same local field factor to construct the kernel, and scaled the results to the positive fermion. Because of the ease of calculations using the simple model, a wide range of densities and mass ratios can be easily investigated. Calculating the different interactions in the electron-positive fermion gas is also straightforward. % The simple model will be validated, amended or rejected when the neglected electron-positive fermion correlation energy is calculated as a function of density.
These interactions and the impact on superconductivity and transport are presented in the following paper.

%%%%%%%%%%55%%%%%%55	Fig 10	%%%%%%%%%%%%%%%%%%%%%%%%%%%%%%%%%%%%%
\begin{figure*}[!ht]
	\centering
    \includegraphics[width=0.95\columnwidth]{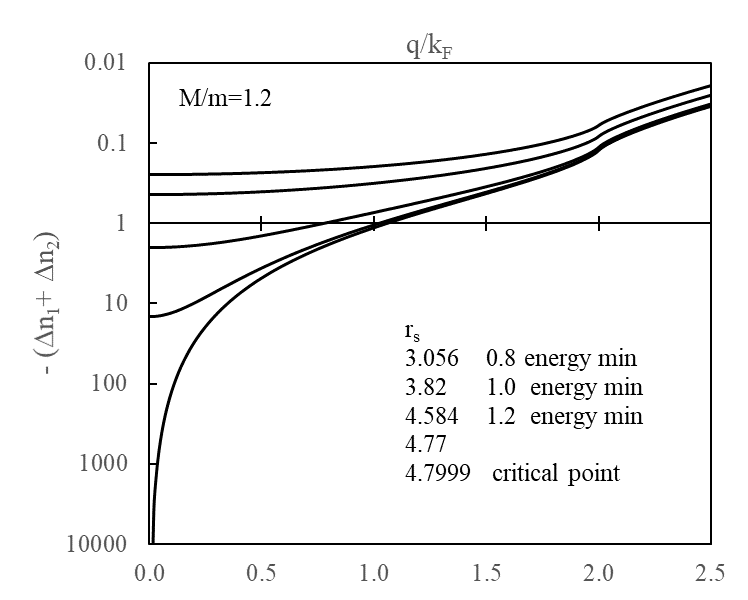}
    \includegraphics[width=0.95\columnwidth]{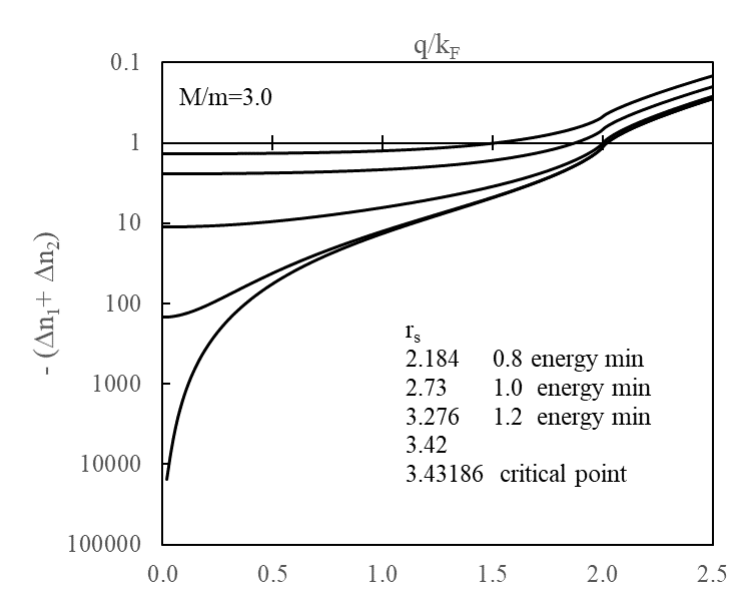}
    \includegraphics[width=0.95\columnwidth]{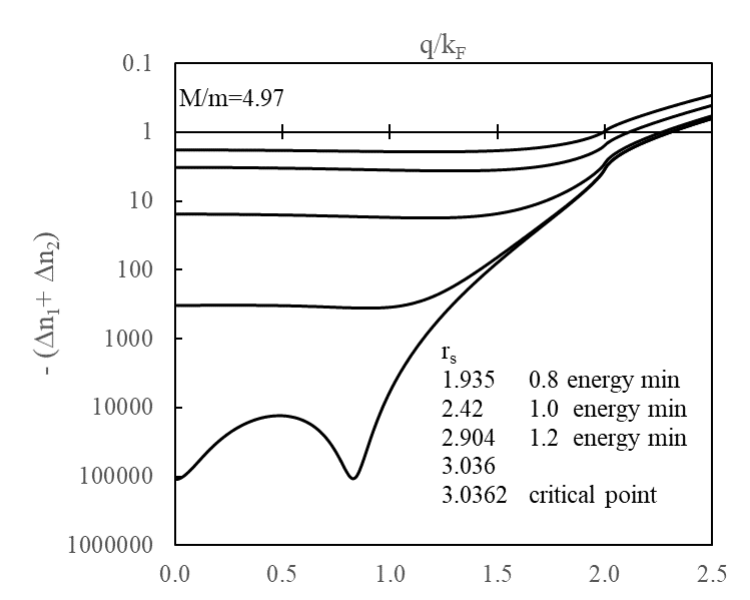}
    \includegraphics[width=0.95\columnwidth]{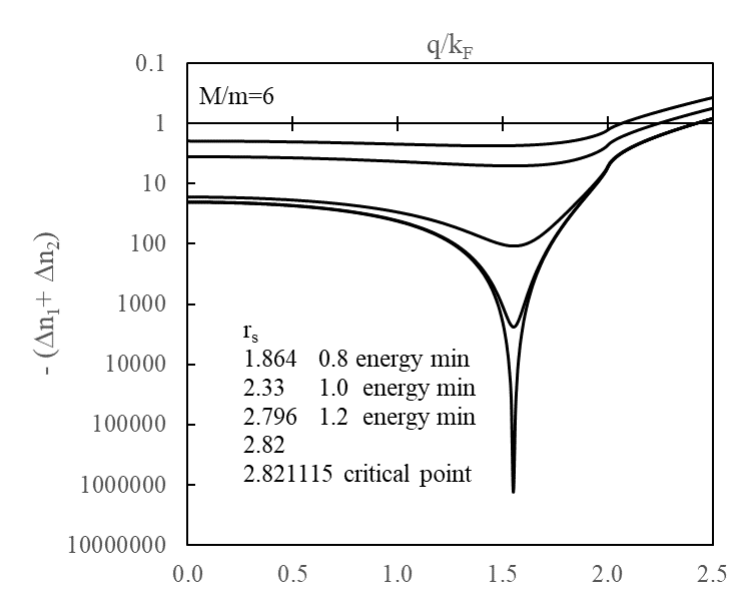}
    \caption{The induced density $\Delta n_1 + \Delta n_2 $ as a function of $q/k_F$ for mass ratios $M/m= 1.2,\,  3.0, \,  4.97, \text{and} \; 
6.0$ for $r_s$ at $0.8,\, 1.0$ and $1.2$ times $r_s$ of the energy minimum, and close to the instability and just before the instability.}
    \label{fig10}
	\end{figure*} 
%%%%%%%%%%%%%%%%%%%%%%%%%%%%%%%%%%%%%%%%%%%%%%%%%%%%%%%%%%%%%%%%%%%%%

%The effect of mass ratio $M/m$ on the sum of the induced densities of the electrons and positive fermions due to an external test charge is shown in Fig. \ref{fig9}. The energy minimum shown in Fig. \ref{fig9}(a), the induced density is substantially lower at finite wave vector starting at $M/m=4.97$.  The effect very close to either instability is much more dramatic as is shown in (b). Below $M/m=4.97$, the instability is at $q/k_F = 0$, at $M/m=4.97$ the triple point is apparent, and above this mass ratio, there are charge density waves. Near the instability, the induced density diverges and a logarithmic scale is needed to show all of the curves on the same graph. Of course, linear response theory is not sufficient near the instabilities and some form of non-linear response theory \cite{ref24} or renormalization is needed.

%%%%%%%%%%%%%%%%%%%%%%%%%%%%%%%%%%%%%%%%%%%%%%%%%%%%%%%%%%%%%%%%%%%%%%%%%%%%%%%%%%%%%%%%%%%%%%%%%%%%%%55
\section{Induced number and charge densities}
%%%%%%%%%%%%%%%%%%%%%%%%%%%%%%%%%%%%%%%%%%%%%%%%%%%%%%%%%%%%%%%%%%%%%%%%%%%%%%%%%%%%%%%%%%%%%%%%%%%%%%%%%%
Consider the electron-positive fermion gas response to a positive test charge. The electrons are attracted to the positive charge and the induced number density $\Delta n_2$ is positive. The positive fermions will be repelled by the test charge and $\Delta n_1$ is negative. The total induced density is $\Delta n = \Delta n_1 + \Delta n_2$. For simplicity I have dropped the subscript $+$ from $\Delta n$. The induced charge density which is relevant to screening is the charge times the induced density, $\Delta \rho_1 = e \Delta n_1$ and $\Delta \rho_2 = - e \Delta n_2$ and $\Delta \rho = e( \Delta n_1 - e\Delta n_2)$. The net coulomb potential from test charge defines the overall dielectric function $1/\varepsilon_{\text{tt}} = 1+ \Delta n_1 - \Delta n_2 $. %Reference \cite{ref7}\red{[7]} Eq. 5.15 states that $1/\varepsilon_{\text{tt}}(q) < 1$ is the condition for stability of the ground state.

Writing out Eq. \eqref{eq7} yields
%%%%%%%%%%%%%%%%%%%%%%%% Ecuacion 17 %%%%%%%%%%%%%%%%%%%%%%%%%%%%%%%%%%%%%%%%%%%%%%%%%%%%%%%%%%%%
\begin{eqnarray}
    V_{2+}^{\text{eff}} = -v \frac{\varepsilon_{\text{ht}} - v \Pi^0_1 (1-2G_{12})}{ \varepsilon_{\text{et}} \varepsilon_{\text{ht}} – v2 \Pi^0_1 \Pi^0_2 (1-2G_{12})^2} \; .    
\end{eqnarray}
%%%%%%%%%%%%%%%%%%%%%%%%%%%%%%%%%%%%%%%%%%%%%%%%%%%%%%%%%%%%%%%%%%%%%%%%%%%%%%%%%%%%%%%%%%%%%%%
When the mass of the positive fermion equals the mass of the electron, the dielectric functions and Lindhard functions become equal and $V_{2+}^{\text{eff}} = -V_{1+}^{\text{eff}}$ and the induced densities given by $\Delta n_{2+} = -\Pi^0_2 V_{2+}^{\text{eff}}$ become equal with opposite sign and the induced number density $\Delta n = \Delta n_1 + \Delta n_2 = 0$ for all values of $q$ and density. The external positive charge attracts the electrons and repels positive fermions in exactly the same amount when the masses are equal. Similarly $\Delta \rho /e = (\Delta n_1 - \Delta n_2) = -2\Delta n_2$. Furthermore, the denominator $\Delta$ becomes factorable into the product
%%%%%%%%%%%%%%%%%%%%%%%% Ecuacion 18 %%%%%%%%%%%%%%%%%%%%%%%%%%%%%%%%%%%%%%%%%%%%%%%%%%%%%%%%%%%%
\begin{eqnarray}
    \Delta = \left(\varepsilon_{\text{et}} – v \Pi^0_2 (1-2G_{12})\right) \left(\varepsilon_{\text{et}} + v \Pi^0_2 (1-2G_{12})\right)\; ,
\end{eqnarray}
%%%%%%%%%%%%%%%%%%%%%%%%%%%%%%%%%%%%%%%%%%%%%%%%%%%%%%%%%%%%%%%%%%%%%%%%%%%%%%%%%%%%%%%%%%%%%%%
and the first factor cancels the numerator. The denominator is rewritten and the resulting effective interaction for M/m=1 is given by
%%%%%%%%%%%%%%%%%%%%%%%% Ecuacion 19 %%%%%%%%%%%%%%%%%%%%%%%%%%%%%%%%%%%%%%%%%%%%%%%%%%%%%%%%%%%%
\begin{eqnarray}\label{eq19}
    V_{2+}^{\text{eff}} = -v/ (1- (2 + (G_{2+} – 2G_{12}))v \Pi^0_2) = -V_{1+}^{\text{eff}}\; .
\end{eqnarray}
%%%%%%%%%%%%%%%%%%%%%%%%%%%%%%%%%%%%%%%%%%%%%%%%%%%%%%%%%%%%%%%%%%%%%%%%%%%%%%%%%%%%%%%%%%%%%%%
At $q=0$, $\Delta n_2 = 1/2 $ and $\Delta n_1 = -1/2 $ and the induced density $\Delta n = (\Delta n_1 + \Delta n_2) = 0$, and the induced charge density $\Delta \rho/e = (\Delta n_1 - \Delta n_2) = -1$.     

The case of equal masses is unusual because in the simple model because $ \Delta n_1 = -B_2/B$ appears to diverge when $B=0$, but since $B = B_1 + B_2$ and $B_1 = B_2$, $\Delta n_1 = B_2/B = 1/2$ and there is no divergence at $q=0$. This is because $B_1$ and $B_2$ in the numerator each separately diverge and are equal in magnitude to $1/2 B$. However, there is a divergence when the masses are not equal and $B_1$ does not equal $B_2$.

The effective potentials Eqs. (\ref{eq7} \& \ref{eq8}) are used to yield the induced densities:
%%%%%%%%%%%%%%%%%%%%%%%% Ecuacion 19 %%%%%%%%%%%%%%%%%%%%%%%%%%%%%%%%%%%%%%%%%%%%%%%%%%%%%%%%%%%%
\begin{eqnarray}
    \Delta n_2 &= & -\Pi^0_2 V_{2+}^{\text{eff}} \\
    &=& q_{TF2}^2 L\left( \frac{q}{k_F} \right) \left( 1 - v \left( G_{1+} - 2G_{12} \right) \right)\frac{\Pi^0_1}{q^2 \Delta} \nonumber
\end{eqnarray}
\begin{eqnarray}
    \Delta n_1 &= & -\Pi^0_1 V_{1+}^{\text{eff}} \\
    & = & q_{TF1}^2 L\left( \frac{q}{k_F} \right) \left( 1 - v \left( G_{2+} - 2G_{12} \right) \right)\frac{\Pi^0_2}{q^2 \Delta} \nonumber
\end{eqnarray}
%%%%%%%%%%%%%%%%%%%%%%%%%%%%%%%%%%%%%%%%%%%%%%%%%%%%%%%%%%%%%%%%%%%%%%%%%%%%%%%%%%%%%%%%%%%%%%%
Where $L(q/k_F)$ is the dimensionless Lindhard function \cite{ref24} that equals $1$ at $q=0$, $1/2$ at $q=2k_F$, and falls off quickly at larger $q$. The denominator was discussed in detail in Section III.

The general solution for a positive point charge disturbance at $q=0$ is given by $\Delta n_{2+} = B_1/B$ and $\Delta n_1 = -B_2/B$, where $B_1 = 1/\kappa_1$ is the bulk modulus (inverse of the compressibility) of the positive fermions in a uniform background and $B = B_1 + B_2$ is the total bulk modulus of the electron-positive fermion gas. In obtaining this result, the inverse mass scaling of the bulk modulus of the noninteracting fermion system $B_0 = 2/3 n \varepsilon_F$ which varies as $1/m$ was used. At $q=0$, the general results are $\Delta n = \Delta n_1 + \Delta n_2 = (B_1 – B_2)/B$ and $\Delta \rho/e = (\Delta n_1 - \Delta n_2) = - (B_1 + B_2)/B = -1$. The induced densities diverge when $B=0$. This is the compressibility instability shown in Fig. \ref{fig1}.
%%%%%%%%%%%%%%%%%%%%%%%%%%%%%%%%%%%%%%%%%%%%%%%%%%%%%%%%%%%%%%%%%%%%%%%%%%%%%%%%%%%%%%%%%%%%%%%%%%%%%%%%%%%%%%%%%%%%%%%%%%%

The effect of mass ratio $M/m$ on the sum of the induced densities of the electrons and positive fermions due to an external test charge is shown in Fig. \ref{fig9}. The energy minimum shown in Fig. \ref{fig9}(a), the induced density is substantially lower at finite wave vector starting at $M/m=4.97$.  The effect very close to either instability is much more dramatic as is shown in (b). Below $M/m=4.97$, the instability is at $q/k_F = 0$, at $M/m=4.97$ the triple point is apparent, and above this mass ratio, there are charge density waves. Near the instability, the induced density diverges and a logarithmic scale is needed to show all of the curves on the same graph. Of course, linear response theory is not sufficient near the instabilities and some form of non-linear response theory \cite{ref24} or renormalization is needed.

To show the sensitivity to $r_s$, I have calculated the induced density at five values of $r_s$ for each mass ratio. The first is at $0.8$ times the $r_s$ of the energy minimum to show the effect of pressure. The second is at the energy minimum and then $1.2$ times the energy minimum. The instability point for $q=0$ transitions is at $1.25$ times the energy minimum. These are shown in Fig. \ref{fig10}; the last two curves are close to the instability and just before the instability

The electron-positive fermion gas with point particles and only coulomb interactions has a unique energy minimum. This is the equilibrium density ($r_s$). To achieve a smaller $r_s$ pressure would need to be applied. To achieve a lower density, larger $r_s$, a pseudo potential or other mechanism is required. Fig. \ref{fig10} shows that the induced density is modestly dependent on $r_s$ until the instability is closely approached, and then becomes extremely sensitive.

%%%%%%%%%%%%%%%%%%%%%%%%%%%%%%%%%%%%%%%%%%%%%%%%%%%%%%%%%%%%%%%%%%%%%%%%%%%%%%%%%%%%%%%%%%%%%%%%%%%%%%55
\section{SUMMARY AND CONCLUSIONS}
%%%%%%%%%%%%%%%%%%%%%%%%%%%%%%%%%%%%%%%%%%%%%%%%%%%%%%%%%%%%%%%%%%%%%%%%%%%%%%%%%%%%%%%%%%%%%%%%%%%%%%%%%%

The simple model of the electron-positive fermion gas in this paper, together with the known accurate local field factors of the electron gas, yields a formula for the location of divergences in the response functions that include an instability at $q=0$ and charge density waves at finite $q$.

The model is simple but the resulting phase diagram is complex. A previous paper in 2019 \cite{ref12} used density functional theory with the local spin density approximation to approach the same problem. The simple theory of this paper is in good agreement with the predictions of the $q=0$ and charge density wave instabilities. The simple theory provides an analytic formula implemented on a laptop that allows detailed examination of the phase diagram, response functions and interactions as a function of $r_s$ and the mass ratio $M/m$.

One simple result from the model is that at the $q=0$ instability where the bulk modulus of the entire system is zero and the compressibility diverges, the bulk modulus of the electrons is positive and the bulk modulus of the heavier positive fermions is negative. Thus it is the electrons that are stabilizing the system. This is in contrast to the usual treatment of the electron gas in a uniform positive background. In that case, when the electron bulk modulus becomes zero, it is assumed that the positive background provides a positive bulk modulus that keeps the system stable.  

The simple model predicts that the electron-positive fermion gas has a unique density $r_s$ where the energy is minimum and the gas would be at equilibrium. At a lower density given by $1.25$ times the $r_s$ of the energy minimum, the bulk modulus of the entire system becomes zero and there is a divergence in the response functions at $q=0$. This indicates a phase transition which may be a condensation of the gas into a smaller volume. These unique densities are a function only of the mass ratio $M/m$. The maximum equilibrium $r_s$ is at $4.19$ when $M/m=1$. This is the same value as for the electron gas in a uniform background. The minimum equilibrium $r_s$ is near $1.6$ when the mass ratio is infinity and the kinetic energy of the positive fermions is zero. If the positive fermion is a proton with mass ratio $M/m=1836$, the equilibrium $r_s= 1.65$ which is close to the value previously found for metallic hydrogen \cite{ref13}. In order to allow the theory to be used at $r_s$ larger than the equilibrium density, a pseudo potential \cite{ref18} would be required as in the uniform electron gas. Pressure could be applied to reduce $r_s$.  

The linear response of the electron-positive fermion gas was calculated in 1974 \cite{ref3} in terms of then partially known local field factors. The local field factors of the uniform electron gas are now well known (but poorly understood from physical intuition) as a function of $q$ from rigorous sum rules and Quantum Monte Carlo calculations (\cite{ref14,ref22} and references therein). The simple model uses these local field factors from the uniform electron gas to quantitatively calculate the divergences at $q=0$ and finite q that define the phase boundaries and the phase diagram.

The results of the linear response theory are that below $M/m=4.97$, there are no charge density waves, and the only instability is the $q=0$ condensation instability at a specific $r_s$ determined by the mass ratio. Above $M/m=4.97$, there is always an $r_s$ that that has an instability at some critical wave vector $q_c$. At $M/m=4.97$ and a very precise $r_s$, simultaneous divergences appear at $q=0$ and $q/k_F=0.83$ indicating a triple point separating phases where the electron-gas is stable and uniform, a condensation at $q=0$, and a charge density wave. The map of these instabilities as a function of $q$ is complex.  At a given $r_s$ there are regions of $q$ where the response functions are positive and other regions where they are negative. This is clearly shown by the mathematics, but an intuitive physical explanation remains elusive. % The nature of the finite $q$ instabilities is not known.
 
The finite $q$ (charge density wave) instabilities depend the additional screening by the positive fermion and crucially on the value of the local field factors at finite $q$. An earlier Hubbard-like version of the local field factors \cite{ref4} that satisfied the compressibility sum rule at $q=0$, but tended to a constant at large $q$, did not predict charge density waves because the value of the local field function was too small near $q=k_F$. The density functional theory calculation \cite{ref12} predicts the same charge density waves, because the local spin density approximation employed a kernel equivalent to using the same local field factor of this paper.

The presence of instabilities enhances the density response, and the induced density is shown to be extremely large when approaching the instabilities. In the following paper, using the Kukkonen-Overhauser \cite{ref4,ref10,ref11} approach, re-derived in Appendices B \& C, is used to calculate the electron-electron, positive fermion-positive fermion and electron-positive fermion interactions and the collective modes. All are affected by the local field factors and the presence of charge density waves. Proximity to an instability significantly enhances superconductivity, leads to a large $T^2$ term in the normal state electrical resistivity and results in a soft collective mode.

Charge density waves are a clear prediction of this simple model (and by the density functional theory calculation), but are they really present in the electron-positive fermion gas? The advantage of the simple model is that the origin of the charge density waves is clear. They occur if $(1+ (1- G_{2+} -v\Pi^0_1/\varepsilon_{\text{ ht}})  v \Pi^0_2) = 0$ (see Eq. \eqref{eq14}). The existence of the charge density wave depends crucially on the terms $G_{2+}$ and $-v\Pi^0_1/\varepsilon_{\text{ht}}$. If either of these terms are smaller, then the charge density wave will be moved out to larger $q$ or disappear altogether.

I believe that the fundamental issue is the intermediate q dependence of the local field factors. Quantum Monte Carlo calculations at $r_s = 1 \, \& \,  2$ show that the approximate $q^2$ dependence continues from $q=0$ to close to $q= 2k_F$ \cite{ref17,ref20}. I have assumed that this $q^2$ behavior continues as $r_s$ becomes larger. %In density functional theory, this is equivalent to the local spin density approximation for the exchange and correlation kernel. 
Previous Quantum Monte Carlo calculations were conducted at $r_s = 2 \, , \,  5 \, \& \, 10$, but there was very little data below $2k_F$ \cite{ref25}. The interpolation formula of Ref. \cite{ref17} is consistent with all of the current Quantum Monte Carlo data. However for $r_s$ greater than $2$, there is no significant Quantum Monte Carlo data between $q=0$ and $q= 2k_F$.

A recent paper \cite{ref26} calculated the local field factor due to exchange only (no correlation) and found that this exchange factor agreed with the Quantum Monte Carlo calculations at $r_s = 1 \,  \& \,  2$, but showed a behavior less than $q^2$ at $r_s = 5$ and $10$.

The charge density wave appears in the approximate range $q=0.8 -2.2 \, k_F$ for mass ratios $> 4.97$, and $r_s < 2.4$. The local field factor for the electrons $G_{2+}$ is most likely accurate for these small values of $r_s$.  However the using the scaling of this paper and Ref. \cite{ref12} yields the effective $r_s$ of the positive fermions to be $M/m$ times that of the electrons. The effective $r_s$ of the positive fermions is greater than $10$ and in the region where no Quantum Monte Carlo calculations of the local factor exist.

The positive fermion local field factor appears in $\varepsilon_{\text{ht}} = 1+v (1 – G_{1+}) \Pi^0_1$, and if $G_{1+}$ is the Hubbard approximation or as in the RPA equals zero, there is no charge density wave. The existence or absence of the charge density wave depends on the electron fermion local field factor at low density $r_s > 10$.

Calculation of the missing electron-positive fermion correlation energy will improve the simple model. Calculation of the local field factor of the electron gas at lower densities will confirm or disprove the presence of charge density waves.

In the electron-positive fermion gas, each fermion species is distinct and cannot exchange. The magnetic response is simply the sum of the response of the electrons and the positive fermions. There are no spin density waves. 

However if the two fermions could exchange (for example two pockets of electrons), an additional exchange term is introduced in the linear response theory which can lead to divergences. The current model of the electron-positive fermion gas has equal numbers of electrons and positive fermions. It is compensated. However there may be situations where they are not equal and the background partially arises from another source. This may lead to other interesting predictions. The simple model can be adapted to these situations. The key ingredients are the second species that provides additional screening and possible exchange with the first species. One could foresee a system that is close to both a charge density wave and spin density wave instability. Application of a magnetic field provides another variable that could end up tuning the proximity to the instability. Of course, the simple model can be readily extended to two dimensions.

No attempt is made to map the simple model onto any real system where comparison with experiment is possible. Previous authors have applied similar theories to the electron-hole liquid in semiconductors, liquid metallic hydrogen, and suggested applications to semi-metals with pockets of electrons and holes \cite{ref2,ref3,ref4,ref7,ref9,ref13}.

%There is a maximum $r_s$ in the linear response theory below which there is no possibility of a charge density wave at any $q$, only a $q=0$ instability. This maximum $r_s$ is smaller than the $r_s$ for the $q=0$ instability for $M/m>4.97$, and below the equilibrium $r_s$ for $M/m<9$. This region where there is no instability is examined in detail. The induced charge density is calculated and shown to be extremely sensitive when approaching any of the instabilities. 
%
%No attempt to directly compare with any experiments is made here, but previous authors have applied similar theories to the electron-hole liquid in semiconductors, liquid metallic hydrogen and suggested application to semi-metals with pockets of electrons and holes \cite{ref2,ref3,ref4,ref7,ref9}. The complexity of the phase diagram invites closer examination. 
%
%The results presented in this paper will be used along with the Kukkonen-Overhauser \cite{ref4,ref10,ref11} approach to calculate the electron-electron, positive fermion-positive fermion and electron-positive fermion interactions and the collective modes of the electron-positive fermion gas in the following paper. Transport properties and superconductivity will also be addressed.

%%%%%%%%%%%%%%%%%%%%%%%%%%%%%%%%%%%%%%%%%%%%%%%%%%%%%%%%%%%%%%%%%%%%%%%%%%%%%%%%%%%%%%%%%%%%%%%%%%%%%%%%%
\section*{Acknowledgment}
%%%%%%%%%%%%%%%%%%%%%%%%%%%%%%%%%%%%%%%%%%%%%%%%%%%%%%%%%%%%%%%%%%%%%%%%%%%%%%%%%%%%%%%%%%%%%%%%%%%%%%%%%%%%%

I appreciate technical discussions with Jan Herbst, thank him for his careful reading of the manuscript. I thank Shiwei Zhang for comments. Daniel Arturo Brito Urbina prepared the publication version of this paper. I am interested in discussions or collaborations to gain a physical understanding of the phase diagram,  calculate the missing electron-positive fermion correlation energy, applications of the simple model to experimental data, investigation of magnetic properties, and in extending the simple model to two dimensions. Comments and criticism are invited.

\appendix
\section{SIMPLE MODEL FOR THE ELECTRON-POSITIVE FERMION GAS}

\subsection{Uniform Electron Gas}
The electron gas in a uniform positive background is the well-studied and textbook example \cite{ref11,ref16} used to calculate many properties of metals. This model has only one parameter, $r_s$, the linear measure of the density of the electron gas $(1/n = 4\pi (r_s a_0)^3/3$ and $a_0$ is the Bohr radius). The uniform positive background deserves further discussion. It is thought of as an average of the lattice of ions in a metal that provides overall charge neutrality for the system. The charge density of the background is viewed as an independent variable that determines the electron gas density characterized by $r_s$. The energy of the background is not considered explicitly and the energy of the system is considered as only the energy of the electrons which has a minimum at $r_s = 4.19$. To achieve higher or lower $r_s$ the background has to do work. To keep the electron gas at a different $r_s$, the background has to be rigid, and some authors use the term rigid uniform positive background. However if the background is rigid, it does not support vibrations and the system would have no phonons. In condensed matter theory, the response of the lattice is treated independently, and the electron gas is assumed able to respond so fast that the electrons are responding to the instantaneous local configuration of the slower moving ions. This is the Born Oppenheimer approximation.

In addition to charge neutrality, the background has another important function. The bulk modulus $B$ of the electron gas, determined by the change in energy with respect to volume, becomes zero at $r_s = 5.25$, and is negative at lower densities (larger $r_s$). The system becomes unstable when $B$ becomes negative. In standard electron gas theory, it is argued that this is not important because the system bulk modulus is the sum of the electron bulk modulus and the background bulk modulus, which is somehow sufficiently positive to keep the uniform electron gas system stable. Pseudo potentials are introduced to model some of the properties of the background (see Ref \cite{ref14} for example).

The most important point for the purpose of this paper is to recognize that the electron gas has a uniform solution for large $r_s$ as long as the background is uniform. The uniform solution may not have the lowest energy at very large $r_s$. 

The electron gas properties have been studied using various calculational methods up to $r_s$ of approximately $120$. A uniform electron density remains a solution within the approximations. A body centered cubic Wigner lattice at $r_s$ somewhat above 100 has lower energy than the uniform gas. Other possibilities include ferromagnetic alignment of the electron spins, charge density waves and spin density waves. At large $r_s$, the different possible phases of the electron gas all have very low and similar energies, and calculations must have great accuracy in order to differentiate potential phases \cite{ref11,ref17}.

The microscopic theory of an electron-positive fermion gas, and the theory of stabilized jellium incorporating pseudopotentials are two existing models that describe the electron gas in a positive neutralizing background. These models allow calculations of the equation of state of the combined system, and thus the bulk modulus which determines the thermodynamic stability \cite{ref18}. The goal of this paper is to gain a physically intuitive understanding from these models, without requiring detailed new calculations.

The approach is to consider a two component degenerate gas with equal numbers of electrons and positive fermions. The positive fermions can be holes in a semiconductor or semimetal that have masses close to the electron mass, or could be protons in metallic hydrogen or ions in a solid or liquid metal. The point fermions have mass $M$, and become classical particles in certain limits. Some previous work concentrated on holes, and the problem was known as the electron hole liquid EHL \cite{ref2,ref3,ref4}. In many of the prior papers, the problem is treated in the center of mass system wich is useful when comparing to excitons and bound states. This paper focuses on the electrons and holes as distinct particles, which aids in developing the physical picture discussed below.

The positive fermions are the neutralizing background for the electrons and vice versa. The electrons and positive fermions each have interactions amongst themselves including Coulomb, exchange and correlation interactions, and the electrons and holes also interact with each other through the Coulomb interaction and this leads to electron-hole correlation. There is no exchange interaction between electrons and positive fermions because they are distinguishable particles. 

Prior work shows that the electron gas has a uniform density solution as long as the background is uniform. The problem of a positive fermion gas is exactly the same problem because equations depend on the square of the charge, and the electron gas results can be used directly with substitution of the fermion mass $M$ for the electron mass $m$. To lowest order, the effect of the positive fermions is to provide a uniform background for the electrons, and vice versa. The remaining part of the interaction, electron-positive fermion correlation, can be treated as a perturbation but is neglected here. Its inclusion will change the numbers slightly, but should not affect the physical picture and overall conclusions. 

With these assumptions, the energy of the system is simply the sum of the energies of the electrons and the positive fermions. The energy per electron is given by
%%%%%%%%%%%%%%%%%%%%%%%%%%%%%%%%%%%5 Equation A1    %%%%%%%%%%%%%%%%%%%%%%%%%%%%%%%%%%%%%%%
\begin{eqnarray}\label{app_1}
    \frac{E}{N} \! &=& \! \frac{2.21}{r_s^2} - \frac{0.916}{r_s} - \frac{0.2846}{1+1.0529 \sqrt{r_s} + 0.3334 r_s}    \; , \nonumber \\
            && \; \; \;  \text{Ry}(e^2/2a_0 =13.6) \text{eV}
\end{eqnarray}
%%%%%%%%%%%%%%%%%%%%%%%%%%%%%%%%%%%%%%%%%%%%%%%%%%%%%%%%%%%%%%%%%%%%%%%%%%%%%%%%%%%%%%%%%%%%%
The first term is the kinetic energy, the second is the exchange energy and the third term is the electron-electron correlation energy approximated by the simple formula of Ref. \cite{ref19}. More accurate versions of the correlation energy are available \cite{ref17}, but their use would make only a small quantitative difference.

For a positive fermion or a different mass electron in a background having a dielectric constant $\varepsilon_B$, the Bohr radius $a_0=\hbar^2/ m e^2$ is modified by the replacements $m \to M$ and $e^2 \to e^2/\varepsilon_B$.  To obtain the results in standard $\text{Ry}= 13.6 eV$, replace $r_s$ by $(M/m)r_s/\varepsilon_B$ and multiply $E/N$ by $M/m \varepsilon_B^2$ \cite{ref11}. This yields
%%%%%%%%%%%%%%%%%%%%%%%%%%%%%%%%%%%5 Equation A2    %%%%%%%%%%%%%%%%%%%%%%%%%%%%%%%%%%%%%%%
\begin{eqnarray}\label{app2}
    \frac{E}{N} &=& \! \frac{2.21}{\frac{M}{m} r_s^2} - \frac{0.916}{\frac{r_s}{\varepsilon_B}} \nonumber \\
    && \;\;\; \; \; \; - \frac{0.2846 \frac{M}{m \varepsilon_B^2}}{1+1.0529 \! \left( \! \frac{M}{m} \frac{r_s}{\varepsilon_B} \! \right)^{\!\frac{1}{2}} + 0.3334 \frac{M}{m} \frac{r_s}{\varepsilon_B} }    \; , \nonumber \\
            && \; \; \;  \text{Ry}(e^2/2a_0) = 13.6 \text{eV} 
\end{eqnarray}
%%%%%%%%%%%%%%%%%%%%%%%%%%%%%%%%%%%%%%%%%%%%%%%%%%%%%%%%%%%%%%%%%%%%%%%%%%%%%%%%%%%%%%%%%%%%%
%%%%%%%%%%55%%%%%%55	Fig A1	%%%%%%%%%%%%%%%%%%%%%%%%%%%%%%%%%%%%%
\begin{figure}
	\centering
    \includegraphics[width=1.0\columnwidth]{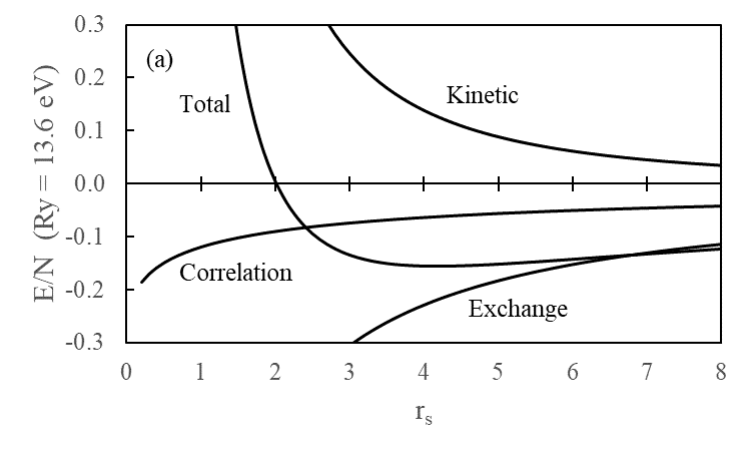}
    \includegraphics[width=1.0\columnwidth]{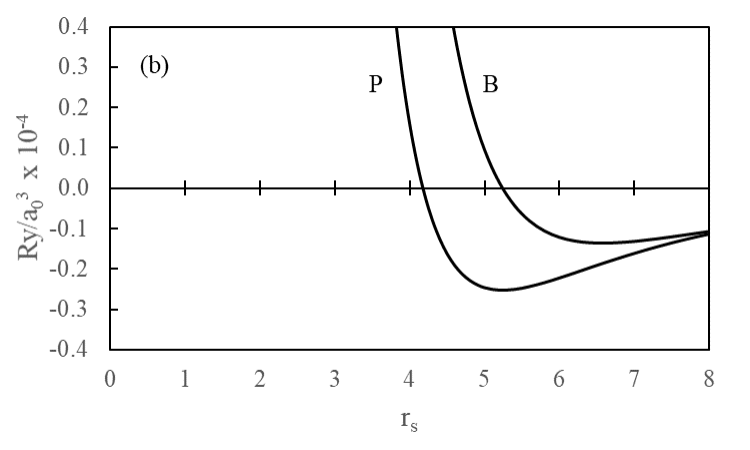}
    \caption{(a) Total energy per electron of the three-dimensional electron gas in a uniform background that is the sum of the kinetic energy, exchange energy and correlation energy contributions. (b) the corresponding pressure and bulk modulus.}
    \label{fig_app1}
	\end{figure} 
%%%%%%%%%%%%%%%%%%%%%%%%%%%%%%%%%%%%%%%%%%%%%%%%%%%%%%%%%%%%%%%%%%%%%
For metallic hydrogen with Coulomb interactions and the proton mass, $\varepsilon_B=1$, $M/m= 1836$, and the effective $r_s$ of the protons is 1836 times that of the electrons. In this case the kinetic energy is negligible and the exchange and correlation energy are dominant. For ions corresponding to other metals, the mass is even larger, and core electrons surrounding the heavier nucleus contribute a background dielectric constant and an additional non-Coulomb interaction. Note that the correlation energy contribution becomes independent of $M/m$ and proportional to $1/r_s$ like the exchange energy when $(M/m)r_s$ becomes very large. This point is discussed in Ref. \cite{ref19}.

 In order to develop an intuitive physical understanding, the individual terms in Eq. \eqref{app_1} are plotted in Fig. \ref{fig_app1} which shows that the electron gas energy has a shallow minimum at $r_s=4.19$. The minimum is not symmetric. At higher densities (smaller $r_s$) the repulsive kinetic energy rises quickly. At larger $r_s$ the energy increases gradually. Once the energy is given, the pressure $P$ exerted by the electron gas is given by minus the partial derivative of energy with respect to volume. The bulk modulus $B$ is given by minus the volume times the partial derivative of the pressure with respect to volume. The compressibility is $ \kappa = 1/\text{B}$. 

The pressure is zero at the energy minimum at $r_s = 4.19$ and the bulk modulus is zero at $r_s = 5.25$. Again note that the curves are not symmetric. They increase rapidly at smaller rs and are small and rise slowly as $r_s$ increases.

It takes very little energy to expand the electron gas beyond its equilibrium value to larger $r_s$, and takes significant energy to compress the gas to higher densities. Numerical calculations show that the electron gas with the uniform background remains uniform to very large $r_s$.  A body centered cubic Wigner crystal has a lower energy than the uniform gas at $r_s=106$, and a ferromagnetic phase has a lower energy at even larger $r_s$. The differences in energy between uniform phase and Wigner crystal and ferromagnetic phase is tiny.  See Refs. \cite{ref11} \& \cite{ref14} and further references therein.

In summary, the electron gas in a uniform background has a minimum energy at $r_s=4.19$. It takes work to either increase or decrease the density. This work must be done by the uniform background.

The electron gas kinetic energy shown in Fig.\ref{fig_app1} has similarities to a classical gas. The classical gas has uniform density; the pressure is positive at all densities, and it takes a box to contain the gas. The electron gas is uniform and has a positive pressure below $r_s=4.19$. In this region, the positive background is acting like a “box” to contain the electron gas that wants to expand. 

However, above $r_s=4.19$, if the background was not rigid and could follow the electron gas without energy cost, the box could expand and the electron gas and the positive background would remain at $r_s=4.19$ and only occupy a portion of the volume of the box. The rest of the box would be empty. This is the crucial difference between a classical gas and the fermion gas-there is an energy minimum.

In order to go above $r_s=4.19$, the electron gas would be under tension. The background has to pull on the electron gas to make it less dense. Equally, the electron gas is pulling on the background. To achieve the low density of Cesium at $r_s=5.63$ would require expansion beyond the point where the electron gas bulk modulus turns negative.  The background would have to provide not only expansion, but also a positive bulk modulus so that the overall system would be stable. To obtain the higher density of lithium, $r_s=3.25$, the background would have to compress the electron gas.

\subsection{Electron-positive fermion gas}

\begin{figure}[h!]
	\centering
    \includegraphics[width=1.0\columnwidth]{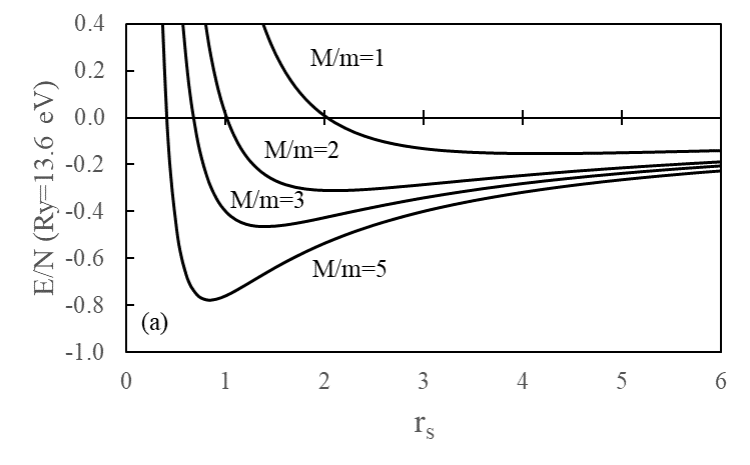}
    \includegraphics[width=1.0\columnwidth]{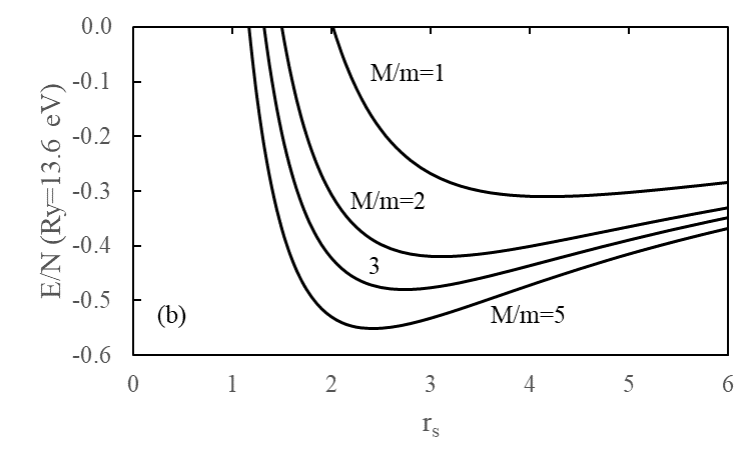}
    \caption{(a) Energy per particle (electron or positive fermion)
of the single Fermi gas with a uniform rigid neutralizing back-
ground plotted versus $r_s$ for different masses, and (b) total
energy of the electron-positive fermion gas which is simply
the sum of the energies of the electron $M/m = 1$ and the
heavier positive fermion.}
    \label{fig_app2}
	\end{figure}

The positive fermions are assumed to have a heavier mass than the electrons. This first term in Eq. \eqref{app2} is the kinetic energy which scales inversely as the mass. For a heavier mass, the kinetic energy is lower. The negative exchange energy is independent of mass and only depends on charge and physical density. The correlation energy is weakly independent on mass. The overall energy becomes lower as the mass increases.  This behavior is shown in Fig. \ref{fig_app2} where the total energy is plotted for $M/m=1, \, 2, \, 3 \,  \& \, 5$.

As the mass ratio is increased, the positive kinetic energy is reduced while the negative exchange is unchanged and negative correlation energy is only weakly affected. Therefore, the energy minimum moves to smaller $r_s$ as the mass ratio increases. The energy minimum becomes deeper and more pronounced as the mass increases. The total energy in the simple model is simply the sum of the energy for $M/m=1$ and the energy at the mass of the positive fermion.

\begin{figure*}[!ht]
	\centering
    \includegraphics[width=1.0\columnwidth]{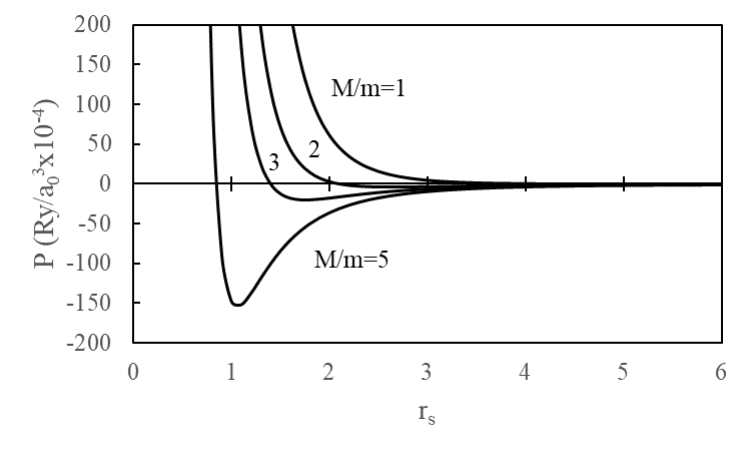}
    \includegraphics[width=1\columnwidth]{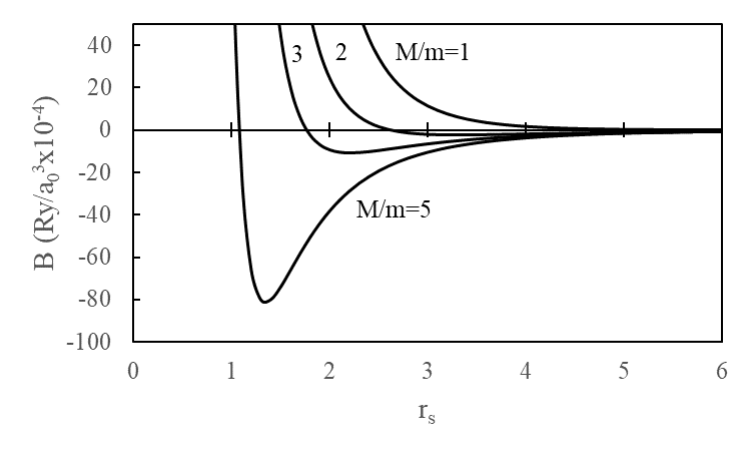}
    \includegraphics[width=1.0\columnwidth]{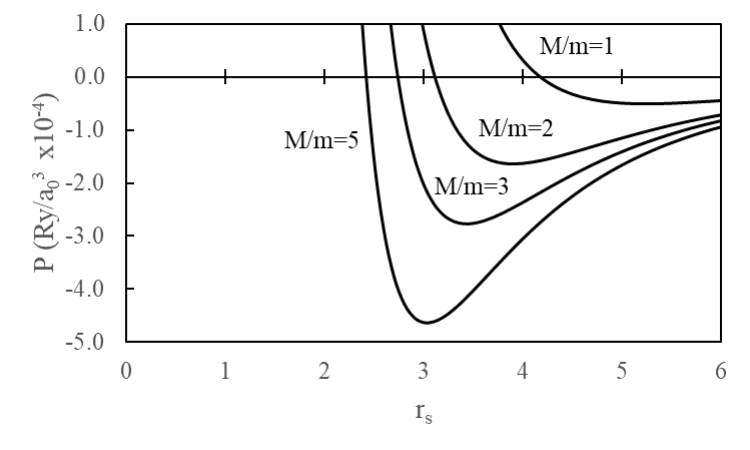}
    \includegraphics[width=1\columnwidth]{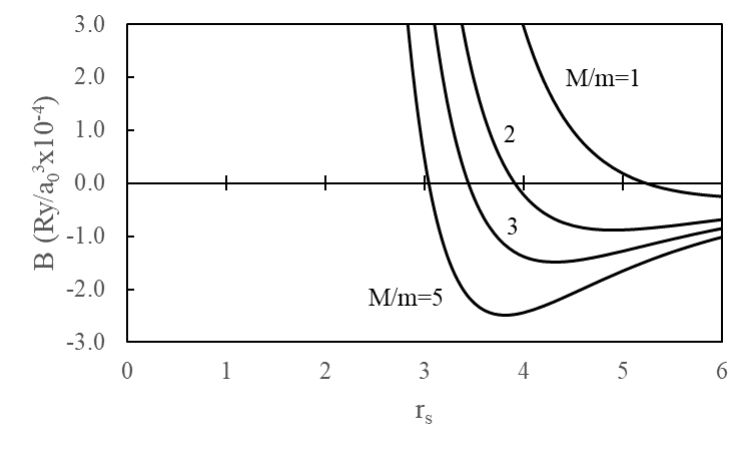}
    \caption{Top - Pressure and bulk modulus of the single species Fermi gas (electron or positive fermion) plotted versus 
$r_s$ for different masses of the particle. Bottom - The total pressure of the electron-positive fermion gas which is the sum of the of the electron gas with M/m=1 and the positive fermion gas with mass M.}
    \label{fig_app3}
	\end{figure*}

The pressure and bulk modulus are just derivatives of the energy with respect to volume. At the energy minimum, the pressure equals zero. At the pressure minimum, the bulk modulus equals zero. These are plotted versus $r_s$ in Figs. \ref{fig_app3} for the same mass ratios as in Fig. \ref{fig_app2}.

%%%%%%%%%%55%%%%%%55	Fig A2	%%%%%%%%%%%%%%%%%%%%%%%%%%%%%%%%%%%%%
%%%%%%%%%%%%%%%%%%%%%%%%%%%%%%%%%%%%%%%%%%%%%%%%%%%%%%%%%%%%%%%%%%%%%

The point where the pressure becomes zero corresponds to the energy minimum.  This is $r_s = 4.19$ for the electron mass and becomes smaller as the mass ratio increases. It is important to note that the point where the pressure of the electron-positive fermion gas is zero, the pressure of the electron gas is positive and the pressure of the positive fermion gas is negative and equal so that they just cancel.

%\newpage
%\begin{minipage}{12cm}
    %%%%%%%%%%55%%%%%%55	Fig A3	%%%%%%%%%%%%%%%%%%%%%%%%%%%%%%%%%%%%%
%%%%%%%%%%%%%%%%%%%%%%%%%%%%%%%%%%%%%%%%%%%%%%%%%%%%%%%%%%%%%%%%%%%%%
%\end{minipage}

% The bulk moduli for the same mass ratios are shown in Fig. \ref{fig_app4}.

%%%%%%%%%%55%%%%%%55	Fig A4	%%%%%%%%%%%%%%%%%%%%%%%%%%%%%%%%%%%%%
%\begin{figure}[h!]
%	\centering
%    \includegraphics[width=1.0\columnwidth]{figures/fig15_1.png}
%    \includegraphics[width=1.0\columnwidth]{figures/fig15_2.png}
%    \caption{(a) Bulk modulus of the Fermi gas (electron or positive fermion) plotted versus $r_s$ for different masses of the particle, and (b) the total bulk modulus of the electron-positive fermion gas which is the sum of the bulk modulus of $M/m=1$ and the bulk modulus of the positive fermion.}
%    \label{fig_app4}
%	\end{figure} 
%%%%%%%%%%%%%%%%%%%%%%%%%%%%%%%%%%%%%%%%%%%%%%%%%%%%%%%%%%%%%%%%%%%%%

As the mass ratio increases, the energy minimum moves to higher density (smaller $r_s$) and becomes deeper. Similarly, the pressure curve becomes steeper and its minimum is deeper. The bulk modulus becomes zero at lower $r_s$, and the slope is much steeper as it crosses zero. 

When the bulk modulus of the electron-positive fermion gas becomes zero, the system is unstable at $q=0$. At this point, the electron gas bulk modulus is positive, and the positive fermion gas bulk modulus is negative. This means that the lighter electrons are stabilizing the gas. For the uniform electron gas, it is assumed that the background is stabilizing the electron gas. Here it is the opposite.

The points where the pressure becomes zero and the bulk modulus becomes zero were calculated as a function of mass ratio and are presented in Section II of the main text.

\subsection{Comparison to previous work}
\begin{figure}[h!]
	\centering
   \includegraphics[width=1.0\columnwidth]{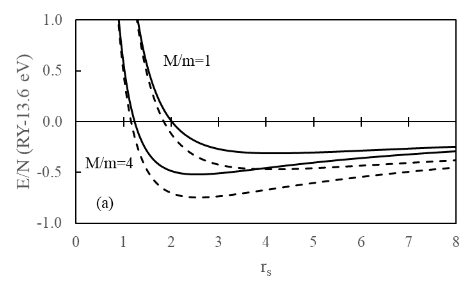}
    \includegraphics[width=1.0\columnwidth]{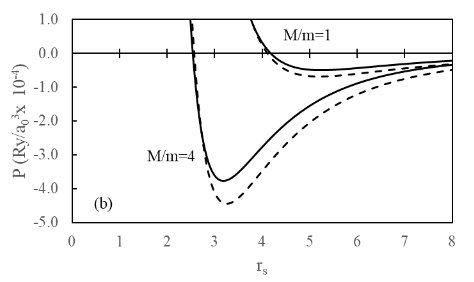}
    \includegraphics[width=1.0\columnwidth]{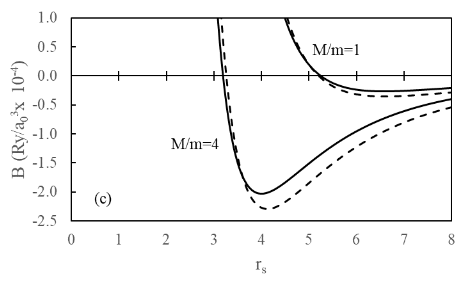}
    \caption{Comparison of the simple model of this paper to universal behavior in electron hole liquid of Ref. \cite{ref20}. The ground state energy is compared in (a), the pressure and bulk modulus obtained by differentiating the ground state energy are shown in (b) and (c).}
    \label{fig_app5}
	\end{figure}

I have compared the energy, pressure and bulk modulus of the simple model to previous work. The electron hole liquid was the subject of significant research in the 1970s \cite{ref2,ref3}. A primary goal was to calculate the correlation energy to predict the overall ground state energy. A short paper by Vashishta and Kalia in 1982 \cite{ref20}  summarized results and gave an analytic formula that fit the calculations in the mass range $M/m=1-4$. The simple model is compared to this analytic formula in Fig. \ref{fig_app5}. 

%%%%%%%%%%55%%%%%%55	Fig A5	%%%%%%%%%%%%%%%%%%%%%%%%%%%%%%%%%%%%%
%%%%%%%%%%%%%%%%%%%%%%%%%%%%%%%%%%%%%%%%%%%%%%%%%%%%%%%%%%%%%%%%%%%%%

The simple expression of Vashishta and Kalia \cite{ref20} modeled a fully self-consistent numerical calculation of the exchange and correlation energy of the electron hole-liquid. This calculation included all correlations. The simple model used in this paper includes electron-electron and positive fermion-positive fermion correlations, but does not include electron-positive fermion correlation energy. These two energies are compared in Fig. \ref{fig_app5}(a) which shows good qualitative agreement, but the self-consistent energy is approximately $0.15$ Ry below the simple curve at $M/m=1$ and $0.22 \,  \text{Ry}$ lower at $M/m=4$. This is likely the additional missing correlation energy. Differentiating both curves with respect to volume shows that the pressure, Fig. \ref{fig_app5}(b), also has a close qualitative agreement with good quantitative agreement near the energy minimum where $P=0$. Differentiating again to get the bulk modulus shows that the bulk modulus is in quite good agreement, and this is the quantity that determines the $q=0$ behavior of the response functions.

%%%%%%%%%%55%%%%%%55	Fig A6	%%%%%%%%%%%%%%%%%%%%%%%%%%%%%%%%%%%%%
\begin{figure}[h!]
	\centering
    \includegraphics[width=1.0\columnwidth]{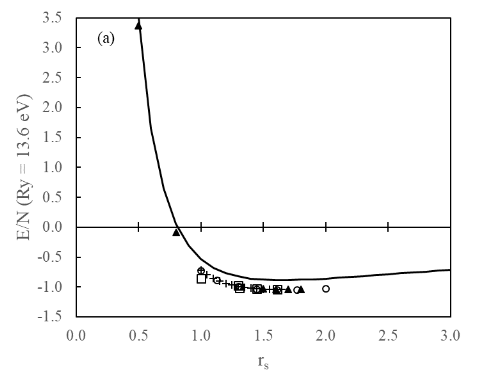}
    \includegraphics[width=1.0\columnwidth]{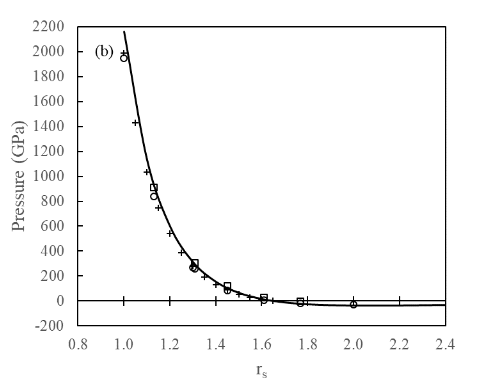}
    \caption{(a) Energy of the electron-proton gas as a function of $r_s$ . The solid curve is from the simple model at $M/m=1836$. The other data are from previous calculations that include the electron-proton correlation energy that is not included in the simple model. The triangles are
from Chakravarty and Ashcroft \cite{ref6} for the gas (liquid) phase. The crosses are from Hammerberg and Ashcroft \cite{ref5} for the protons on an fcc lattice. The circles and squares are from Ceperley and Alder \cite{ref8}, circles for the protons on a static fcc lattice, and squares for a dynamic lattice. (b). Pressure of the electron-proton gas as a function of $r_s$ . The solid curve is the result of differentiating the energy of the simple model with respect to volume. The other data are from previous calculations. The crosses are from \cite{ref5} for the protons on an fcc lattice. The circles and squares are from \cite{ref8}, circles for the protons on a static fcc lattice, and squares for a dynamic lattice.}
    \label{fig_app6}
	\end{figure} 
%%%%%%%%%%%%%%%%%%%%%%%%%%%%%%%%%%%%%%%%%%%%%%%%%%%%%%%%%%%%%%%%%%%%%

Figure \ref{fig_app6}(a) shows that the simple approximation of this paper (solid line) does a good job of representing the shape of the energy curve and has a minimum at $r_s= 1.65$ which is close to the value of $1.6$ used by Ashcroft \cite{ref13}. The energy minimum is $-0.885 \, \text{Ry}$ compared to approximately $-1.05 \,  \text{ Ry }$ in previous work. This difference of $0.165 \, \text{Ry}$ represents the contribution of the neglected portion of the electron-proton correlation energy. For comparison, the electron-electron correlation energy is approximately $-0.1 \, \text{Ry}$ at the energy minimum, and the proton-proton correlation energy is approximately $-0.5 \, \text{Ry}$. The energy difference found in previous work between the gas (liquid) and different lattice structures for the protons is typically on the order of $0.005 \, \text{Ry}$. The binding energy of the hydrogen atom is $1 \,  \text{Ry}$ (binding energy conventionally taken to be positive).

 Chakravarty and Ashcroft \cite{ref6} found that the ground state of metallic hydrogen has a metastable point at $r_s=1.64$, with a small difference between the liquid and crystal states. Chakraborty, Kallio, Lantto and Pietilainen \cite{ref7} studied liquid metallic hydrogen as a two-component Fermi fluid and remarked that “the effect of replacing the rigid background by the proton fluid appears to be surprisingly small.” 

Figure \ref{fig_app6}(b) shows close agreement of the calculated pressures of the simple model of this paper and previous work. Since the pressure is the derivative of the energy curve, this implies that the electron-proton correlation energy is relatively insensitive to $r_s$, particularly near the energy minimum, and thus makes little contribution to the pressure curve.

It follows therefore that the bulk modulus, which is proportional to the derivative of the pressure curve with respect to volume, will also have little contribution from the electron-proton correlation energy and that the simple results from this paper should be accurate.

The Fermi wave vector $k_F$ depends only on the density and is given by $3.63/r_s$ in inverse angstroms. The Fermi wave vectors for electrons and holes are equal. The Fermi energy, however, depends inversely on mass: $\varepsilon_F = \hbar^2 k_F^2/2m = m/M \times 50.1/r_s^2 \, \text{eV}$. The corresponding Fermi temperature is $T_F= m/M \times 581,200/r_s^2$. K The plasma frequency goes as $m^{-1/2}$ and the corresponding plasma temperature $T_P = (m/M)^{1/2} \times r_s^{-3/2} \times 547,000 \text{K}$.

There are two distinct regions of interest for the electron-positive fermion gas. The first is when the positive fermions (holes) are nearly the same mass as the electrons.  The second is when the positive fermion is a proton or ion. The Fermi and plasma temperatures for several examples are given in Table \ref{table1}.  

%%%%%%%%%%%%%%%%%%%%%%%%%%%%%%%%%%%%%%5 Tabla 1         %%%%%%%%%%%%%%%%%%%%%%%%%%%%%%%%%%%%%%%%%%%55
\begin{table}[h!]
    \centering
    \begin{tabular}{|c|c|c|c|c|}
       \hline  {\bf Description }   &   $  M/m   $     &    $ r_{s\text{min}}$   &  $T_F(K)$    &        $T_P(K)$      \\
       \hline      Electron     &               1           &               $4.19$  &           $33105$     &    $63777$        \\
        \hline     Hole        &    5      &               2.42        &        19848      &        64980      \\
        \hline     Proton          &        1836           &               1.65            &       116         &       6023            \\
        \hline      Lithium     &        12744   &        1.63       &        17         &            2328       \\        \hline
    \end{tabular}
    \caption{Temperatures equivalent to the Fermi energy and plasma frequency of electrons and posiive fermions with different mass ratios at $r_s$ of the energy minimum.}
    \label{table1}
\end{table}
%%%%%%%%%%%%%%%%%%%%%%%%%%%%%%%%%%%%%%%%%%%%%%%%%%%%%%%%%%%%%%%%%%%%%%%%%%%%%%%%%%%%%%%%%%%%%%%%%%%%%

Holes with mass near the electron mass will be a degenerate Fermi gas at room temperature. However, a proton or a charge with the mass of lithium would only be degenerate at low temperatures. Similarly, the plasma temperatures (frequencies) of protons and ions are much lower than that of the electrons. This means that the electrons are able to respond much faster than the proton or ion.  This justifies the use of the adiabatic or Born Oppenheimer approximation that assumes that the electrons respond to the ions as if they were static. This allows separate treatments of the electron and ion coordinates which is a fundamental assumption of density functional theory. However, if the holes are close in mass to the electron, the adiabatic approximation is not valid.

\section{ RESPONSE FUNCTIONS AND INTERACTIONS OF THE UNIFORM ELECTRON GAS }

This appendix summarizes previous results from the literature and introduces the notation used in this paper. It is presented to allow nonexperts to understand the equations and simple quantum physics involved. The discussion here is for three-dimensions at $T=0$, but can be directly modified for two or one dimension.

The response functions and effective interactions of the uniform electron gas have been studied for almost as long as the ground state energy. Both represent difficult many body problems. The approach here follows the self-consistent mean field approach used by Kukkonen and Overhauser \cite{ref10}. The same results were obtained using Feynman diagrams by Kukkonen and Wilkins \cite{ref15}, and Vignale and Singwi \cite{ref21}. Both approaches are well described in the textbook by Guiliani and Vignale \cite{ref11}. Earlier citations can be found in these references.

The perturbation of the electron gas can be a test charge with just a coulomb interaction which affects both spin electrons equally, or it can be a magnetic field that acts differently on spin up and spin downs. The perturbation can also be another electron that has both charge and spin.
The response of the noninteracting electron gas was calculated by Lindhard in 1954, and the response function is known as the Lindhard function $\Pi^0(q,\omega)$. This is discussed extensively in Ref. \cite{ref11} where they denote the Lindhard function by $\chi_0=-\Pi^0$.

The intractable many-body calculation of the response functions can be reduced to a set of coupled algebraic equations by assuming that the interactions between electrons in the Fermi sea are local and only dependent on the relative spin of the two electrons and the momentum transfer of the interaction. 

Electrons with spin up or down are perturbed by external potentials $V_\uparrow^{\text{ext}}$ and $V_\downarrow^{\text{ext}}$. The wave vector and frequency dependence will not be explicitly shown unless it is necessary to understand the physics. If the external potential is simply a test charge with coulomb interactions, $v= 4\pi e^2/q2$, the effect on spin up and spin down electrons is the same.

The external potentials induce changes in the spin-up and spin-down electron densities $\Delta n_\uparrow$ and $\Delta n_\downarrow↓$.  The effective potential felt by an average spin-up electron in the Fermi sea is
\begin{eqnarray}
    V_\uparrow^{\text{eff}} &=& V_{\uparrow}^{\text{ext}} \! + v \Delta n_\uparrow \! - v G_{\uparrow \uparrow} \Delta n_{\uparrow} \! + v \Delta n_\downarrow \! - v G_{\uparrow \downarrow} \Delta n_\downarrow   \nonumber       \\
    &=& V_\uparrow^{\text{ext}}  \! + v \Delta n_\uparrow (1-G_{\uparrow \uparrow}) \! + v \Delta n_\downarrow (1-G_{\uparrow \downarrow})
\end{eqnarray}
It is important to distinguish the effective potential seen by an average spin up or down electron in the Fermi sea (which determines the density response) from the interaction between specific electrons.

The local field factor $G_{\uparrow \uparrow}$ represents the additional local interaction that a spin-up electron feels due to exchange and correlation interactions with the spin-up induced electron density. Similarly, $G_{\uparrow \downarrow}$ represents the additional correlation interaction that the spin-up electron has with the spin-down induced electron density. The effective potential felt by an average spin-down electron in the Fermi sea is given by
\begin{eqnarray}
    V_\downarrow^{\text{eff}} &=& V_\downarrow^{\text{ect}} \! + v \Delta n_{\uparrow} (1-G_{\downarrow \uparrow}) \! + v \Delta n_{\downarrow} (1-G_{\downarrow \downarrow}) \;  . 
\end{eqnarray}
A paramagnetic electron gas is considered where $G_{\uparrow \uparrow} = G_{\downarrow \downarrow}$ and $G_{\downarrow \uparrow} = G_{\uparrow \downarrow}$. 

The approach is to self-consistently solve for the induced densities in terms of the local field factors. It is the Pauli exclusion principle between parallel electrons, and the coulomb interaction between all electrons that lead to the local field factors. In the random phase approximation, all of the local field factors equal zero.

With a similar substitution for the down spin electrons, the equations can easily be solved. It is simplest and leads to a better intuition to add and subtract the two equations.
\begin{eqnarray}
    V_\uparrow^{\text{eff}} + V_\downarrow^{\text{eff}} &=& V_\uparrow^{\text{ext}} + V_\downarrow^{\text{ext}} \\
        &&  \;  \;  \;  + 2v (\Delta n_\uparrow + \Delta n_\downarrow ) (1 - (G_{\uparrow \uparrow} + G_{\uparrow \downarrow}) ) \nonumber  \\
    V_\uparrow^{\text{eff}} - V_\downarrow^{\text{eff}} &=& V_\uparrow^{\text{ext}} - V_\downarrow^{\text{ext}} \\
        &&  \;  \;  \;  + 2v (\Delta n_\uparrow - \Delta n_\downarrow ) (G_{\uparrow \uparrow} - G_{\uparrow \downarrow} ) \nonumber
\end{eqnarray}
The sum of the two effective potentials only depends on the sum of the induced densities and the sum of the two local field factors. The difference of the two equations only depends on the difference of the induced densities and the difference of the two field factors.

Since the equations only involve sums and differences, it is convenient to introduce new variables, $V_+^{\text{eff}} = (V_\uparrow^{\text{eff}} + V_\downarrow^{\text{eff}})/2$, $V_+^{\text{ext}} = (V_\uparrow^{\text{ext}} + V_\downarrow^{\text{ext}})/2$, $\Delta n_+ = \Delta n_\uparrow + \Delta n_\downarrow$, $G_+ = G_{\uparrow \uparrow}+ G_{\uparrow \downarrow}$  with corresponding notation for the difference equation.

In the new notation, the equations become
\begin{eqnarray}
    V_+^{\text{eff}} &=& V_+^{\text{ext}} + v(1-G_+)\Delta n_+                \\
    V_-^{\text{eff}}    &=& V_-^{\text{ext}} + v(-G_-)\Delta n_-
\end{eqnarray}
The electrons are feeling the effective potential, and first-order perturbation theory shows that the induced density is linearly related to the effective potential by the Lindhard function,  $ \Delta n_\uparrow = -\Pi^0 V_\uparrow^{\text{eff}}/2 $.The factor of two represents that half of the electrons have spin up in a non-polarized electron gas. The same applies to spin-down electrons. With the definitions above $\Delta n_+ = - \Pi^0 V_+^{\text{eff}}$ and $\Delta n_- = -\Pi^0 V_-^{\text{eff}}$.

The standard results are obtained, 
\begin{eqnarray}
    V_+^{\text{eff}} &=& \frac{V_+^{\text{ext}}}{1 + v (1-G_+)\Pi^0} \; = \; \frac{V_+^{\text{ext}}}{\varepsilon_{\text{et}}}
\end{eqnarray}
where $\varepsilon_{\text{et}}$ is defined as the electron test charge dielectric function, and
\begin{eqnarray}
    V_-^{\text{eff}} &=& \frac{V_-^{\text{ext}}}{1-v G_-\Pi^0}
\end{eqnarray}
The test charge - test charge dielectric function $\varepsilon_{\text{tt}}$ is the screened coulomb interaction due to a test charge in the electron gas. Both spin-up and spin-down electrons feel the same external potential.
\begin{eqnarray}
    V_{\text{tt}} &=& \frac{v}{\varepsilon_{\text{tt}}} \;=\; v(1+\Delta n_+) \; = \; v(1-\Pi^0 V_+^{\text{eff}})   \\
    &=& v \left( 1- \frac{v \Pi^0}{1+v(1-G_+)\Pi^0}  \right) \;=\;  \frac{v( 1-vG_+ \Pi^0)}{1+v(1-G_+) \Pi^0} \nonumber
\end{eqnarray}
At this point it is interesting to note that the vertex function derived in the Feynman diagram approach is $\Lambda = 1/(1-G_+ \Pi^0)$ and therefore $1/\varepsilon_{\text{et}} = \Lambda/\varepsilon_{\text{tt}}$. This latter equality is useful to remember because results calculated at different ways may look different but are actually the same.

The most common external potential that acts differently on spin-up and spin-down electrons is a potential caused by the interaction of the spin with an external magnetic field $H$.

A magnetic field splits the degenerate spin-up and spin-down states. A spin-up electron feels an external potential $+\mu_B H$, while a spin-down electron feels $- \mu_B H$. In this case the sum of the external potentials is zero and the difference is $2\mu_B H$.

The magnetic susceptibility is defined as 
\begin{eqnarray}
    \chi &=& \mu_B \frac{\Delta n}{H} \;=\; \frac{ \mu_B^2 \Pi^0}{1-vG - \Pi^0}
\end{eqnarray}
where the susceptibility of the noninteracting electron gas is $\mu_B^2 \Pi^0 $ and $1/(1 - v G- \Pi^0)$ is the susceptibility enhancement due to exchange and correlation.

These local field factors must satisfy many constraints at small and large wave vector. Quantum Monte Carlo calculations have been performed and provide the behavior at intermediate wave vector. For practical applications, the local field factors of the uniform electron gas are sufficiently well known.

The remaining interaction in the uniform electron gas is the interaction between two electrons. In this case, one specific electron in the Fermi sea is considered the perturbation that another specific electron feels. The interactions between parallel and antiparallel electrons are different. It is crucial to distinguish between the specific electrons that are interacting for the average electrons that make up the induced densities from $\Delta n_\uparrow $ and $\Delta n_\downarrow$.

The first issue is to understand what the disturbance or “external potential” seen by the average electrons in the Fermi sea. When we consider a test charge as the external potential, it is viewed as a charge at a fixed position that interacts with the Fermi sea only through the coulomb potential. This implies that the test charge has infinite mass and does not react in any way to the electrons in the Fermi sea. An electron has the same coulomb interaction as a test charge. However this specific electron is identical to other same spin electrons and also has finite mass. This leads to exchange and correlation interactions with the average electrons of the same spin, and a correlation interaction with electrons of opposite spin.
The relative spin of the perturbing electron and the average electron in the Fermi sea must now both be specified which leads to a slightly different notation.

$V_{\uparrow \uparrow}^{\text{ext}} = v \rho_\uparrow (1- 2G_{\uparrow \uparrow})$ is the external potential felt by an average spin up electron in the Fermi sea due to the perturbation by a spin up electron. $V_{\downarrow \uparrow}^{\text{ext}} = v \rho_\uparrow (1- 2G_{\downarrow \uparrow})$ is the external potential felt by an average spin down electron in the Fermi sea due to the perturbation of a spin up electron.

These external potentials can be used in the equations above, but both spins must now be specified.
\begin{eqnarray}
    V_{\text{ee}+}^{\text{eff}} &=& \frac{v \rho_\uparrow(1-2G_{\uparrow \uparrow}) + v \rho_\uparrow(1-2G_{\downarrow \uparrow}) }{2(1+v(1-G_+)\Pi^0)} \\
    &=& \frac{v\rho_\uparrow (1-G_+)}{1+v(1-G_+)\Pi^0} \; = \; \frac{v \rho_\uparrow (1-G_+)}{\varepsilon_{\text{et}}}  \nonumber\\
    V_{\text{ee}-}^{\text{eff}} &=& \frac{v \rho_\uparrow(1-2G_{\uparrow \uparrow}) + v \rho_\uparrow(1-2G_{\downarrow \uparrow}) }{2(1-vG_-\Pi^0)} \\
    &=& -\frac{v\rho_\uparrow G_-}{1-v G_- \Pi^0}    \nonumber   
\end{eqnarray}
These equations for the average potentials felt by an electron in the Fermi sea are exactly the same as the sum and difference of Eqs. (32) \& (33) of Ref. \cite{ref10}, and Eqs. (5.146) \& and (5.147) of Ref. \cite{ref11}.  

In order to calculate electron-electron matrix elements that are needed for superconductivity and the electron-electron scattering contribution to the electrical and thermal resistivities, another step is needed. The wave function for two electrons of the same spin must be antisymmetric. This incorporates exchange in the correct way. The average effective potential between two parallel electrons in the Fermi sea already includes exchange and correlation in an average (local mean field) way. Similarly the effective potential between two opposite spin electrons includes correlation. Using these potentials for matrix elements will double count exchange and correlation between these two specific electrons. 

%%%%%%% Algo pasa aca
The effective potentials for matrix elements and scattering amplitudes are obtained by removing the direct exchange and correlation between parallel spin electrons, and removing the direct correlation between opposite spin electrons. The term $-v \rho_\uparrow 2 G_{\uparrow \uparrow}$ is subtracted from $V_{\uparrow \uparrow}^{\text{eff}}$ and $-v \rho_\uparrow 2G_{\downarrow \uparrow}$ from $V_{\downarrow \uparrow}^{\text{eff}}$.

The effective potentials for matrix elements of parallel and antiparallel electrons are denoted by $W_{\uparrow \uparrow}$ and $W_{\downarrow \uparrow}$. These are to be distinguished from the scattering amplitudes as discussed in Ref. \cite{ref11} page 235. There are several different but completely equivalent ways to write these in terms of the local field factors. I will explicitly show three different ways and how each contributes to a physical interpretation.
\begin{eqnarray}
    W_{\uparrow \uparrow} \!\! &=& \!\! v \rho_\uparrow \! \left( \frac{\Lambda^2}{\varepsilon_{\text{tt}}} - \frac{G_+^2 v \Pi^0}{1-vG_+\Pi^0}                    - \frac{G_-^2 v \Pi^0}{1 - v G - \Pi^0}  \right)   \nonumber\\
    \\
    W_{\downarrow \uparrow} \!\! &=& \!\! v \rho_\uparrow \! \left( \frac{\Lambda^2}{\varepsilon_{\text{tt}}} - \frac{G_+^2 v \Pi^0}{1-vG_+\Pi^0}                    + \frac{G_-^2 v \Pi^0}{1 - v G - \Pi^0}  \right) \nonumber\\
\end{eqnarray}
This way of writing the interactions is connected to Feynman diagrams. The first term is a vertex correction for each electron divided by the test charge test charge dielectric function. This term is obtained by the summing the direct diagrams Refs. \cite{ref15,ref25}. The next two terms involve summing the ladder diagrams. The second term is from ladder diagrams with the same spin. The third term is from ladder diagrams of opposite spin. Both the first term and the second term diverge at $q=0$ when the compressibility sum rule diverges. However the divergences in the first two terms exactly cancel and their sum is not divergent. This is clearly seen by combining the first two terms which yields
\begin{eqnarray}
    W_{\uparrow \uparrow} &=& v \rho_\uparrow \left( \frac{1+(1-G_+)G_+ v \Pi^0}{1+v(1-G_+)\Pi^0} - \frac{G_-^2 v \Pi^0}{1-v G_- \Pi^0}          \right)  \nonumber\\
       \label{B15} \\
    W_{\downarrow \uparrow} &=& v \rho_\uparrow   \left( \frac{1+(1-G_+)G_+ v \Pi^0}{1+v(1-G_+)\Pi^0} + \frac{G_-^2 v \Pi^0}{1-v G_-\Pi^0}     \right)    \nonumber\\
\end{eqnarray}
The denominator of the first term is simply the electron test charge dielectric function $\varepsilon_{\text{et}}$ which is not divergent at the compressibility divergence.

In Ref. \cite{ref11}, Eqs. (5.148) \& (5.149) can be written using this notation as
\begin{eqnarray}
    W_{\uparrow \uparrow} &=& \rho_\uparrow   \left( v - \frac{v(1- G_+)^2 \Pi^0}{1+v(1-G_+)\Pi^0} - \frac{(v G_-)^2 \Pi^0}{1-v G_- \Pi^0}    \right) \nonumber\\
    \\ % \frac{-v(1-G_+)^2 \Pi^0}{1+v (1-G_+)\Pi^0}
    W_{\downarrow \uparrow} &=& \rho_\uparrow   \left( v - \frac{v(1- G_+)^2 \Pi^0}{1+v(1-G_+)\Pi^0} + \frac{(v G_-)^2 \Pi^0}{1-v G_- \Pi^0}    \right) \nonumber\\
    % \frac{-v(1-G_+)^2 \Pi^0}{1+v (1-G_+)\Pi^0}
\end{eqnarray}
% Aqui se acaba el reemplazo
This way of presenting the interaction emphasizes the bare interaction and the sources of screening. Both the first term and second term are divergent at $q=0$, and the divergences cancel.

All three ways of presenting the electron-electron scattering potentials are exactly equal.

In the spin invariant notation using the Pauli matrices, the effective interaction between two electrons have having spins $\sigma_1$ and $\sigma_2$ is given by
\begin{eqnarray}
    V^{\text{eff}}_{\sigma_1 \sigma_2} &=& \frac{W_{\uparrow \uparrow}  + W_{\uparrow \downarrow}}{2} + \sigma_1 \cdot \sigma_2 \frac{W_{\uparrow \uparrow} - W_{\uparrow \downarrow}}{2}
\end{eqnarray}

Since the static local field factors for the uniform gas are known for practical calculations, all of the response functions for the paramagnetic uniform electron gas are completely specified. These interactions only depend on the electron gas density rs and are compared at two different densities $r_s = 2 \, \& \, 5$ which are representative of high density and low density of metals in Ref. \cite{ref22}.

In summary, the uniform electron gas is well documented. With a rigid uniform background, there are no phonons and therefore no BCS superconductivity. With the known local field factors for the paramagnetic gas, there are no density instabilities such as the Mott transition, charge density waves or magnetic transitions or spin density waves for any $r_s < 100$.

\section{RESPONSE FUNCTIONS AND INTERACTIONS OF THE ELECTRON-POSITIVE FERMION GAS.}

In this appendix, the earlier results for the two-component Fermi plasma \cite{ref3,ref4,ref10} are re-derived. In addition to the density, the second species introduces a new parameter-- the relative mass ratio of the two fermions. The response functions from this model have divergences that indicate real phase transitions and charge density waves. BCS-like superconductivity was predicted \cite{ref4,ref9}. The reasons for the re-derivation in this Appendix are to provide an intuitive derivation all in one place for the nonexpert, and to establish the nomenclature used in the main body of the paper. The nonexpert should read Appendix B before this Appendix C.

The positive fermions can be holes with masses close to the electron mass, or heavier particles such as protons or ions. The mass ratio is crucial. For light masses, the Born Oppenheimer approximation does not hold.

Following Ref. \cite{ref3}, the positive fermions have the subscript 1, and the electrons subscript 2. Note that the second fermion does not have to be a positive charge, and the approach will be the same as two species of electrons in a uniform positive background. This will be considered in a future publication  when a second species of electrons with subscript 3 is included.

To allow for non-coulomb interactions between the positive fermions and between positive fermions and electrons we define the bare interaction between two ions as $V_{11}^{\text{b}}$ and the bare interaction between an electron and a positive fermion as $V_{12}^{\text{b}} = V_{21}^{\text{b}}$. For simple electrons and holes, both would be the coulomb interaction $v=4\pi e^2/q^2$ with the appropriate sign for equal or opposite charges. The electron-positive fermion gas is assumed to be paramagnetic and unpolarized. The electrons and positive fermions are distinguishable particles and cannot exchange with each other.

Consider an external perturbation that can act differently on the electrons or positive fermions and differently on spin-up or spin-down particles. The external potential acting on the spin-up positive fermion is denoted as $V_{1\uparrow}^{\text{ext}}$, and the effective potential seen by positive fermion with spin-down is denoted as $V_{1\downarrow}^{\text{eff}}$. These external potentials induce charge densities in the spin-up and spin-down electrons and positive fermions. 
\begin{eqnarray}
    V_{1\uparrow}^{\text{eff}} &=& V_{1\uparrow}^{\text{ext}} + V_{11}^{\text{b}}(1-2G_{1 \uparrow \uparrow}) \Delta n_{1 \uparrow}  \nonumber\\
     && \; + V_{11}^{\text{b}}(1-2G_{1\uparrow \downarrow}) \Delta n_{1\downarrow} \\
     && \; \; + V_{12}^{\text{b}}(1-2G_{12})(\Delta n_{2\uparrow} + \Delta n_{2\downarrow}) \nonumber   \\
      V_{1\downarrow}^{\text{eff}} &=& V_{1\downarrow}^{\text{ext}} + V_{11}^{\text{b}}(1-2G_{1 \downarrow \uparrow}) \Delta n_{1 \uparrow}  \nonumber\\
     && \; + V_{11}^{\text{b}}(1-2G_{1\downarrow \downarrow}) \Delta n_{1\downarrow} \\
     && \; \; + V_{12}^{\text{b}}(1-2G_{12})(\Delta n_{2\uparrow} + \Delta n_{2\downarrow}) \; ,\nonumber \\ 
      \text{\small{where }} G_{12} &=& G_{21}\; .   \nonumber
\end{eqnarray}
There are corresponding equations for the electron effective interactions.  Adding and subtracting as was done for the uniform electron gas in Appendix B, the following coupled equations are obtained with the simplified notation, $V_{1+}^{\text{eff}} = (V_{1\uparrow}^{\text{eff}} + V_{1\downarrow}^{\text{eff}})/2$, $V_{1+}^{\text{ext}} = (V_{1\uparrow}^{\text{ext}} + V_{1\downarrow}^{\text{ext}})/2$, $G_{1+} = G_{1\uparrow \uparrow} + G_{1\uparrow \downarrow}$ and $G_{1-} = G_{1\uparrow \uparrow} - G_{1\uparrow \downarrow}$.
\begin{eqnarray}
    V_{1+}^{\text{eff}} &=& V_{1+}^{\text{ext}} + V_{11}^{\text{b}}(1-G_{1+}) \Delta n_{1+} \nonumber   \\
    && \; + V_{12}^{\text{b}}(1-2G_{12}) \Delta n_{2+}  \label{c3}\\
    V_{1-}^{\text{eff}} &=& V_{1-}^{\text{ext}} + V_{11}^{\text{b}}(-G_{1-}) \Delta n_{1-}    \\
    V_{2+}^{\text{eff}} &=& V_{2+}^{\text{ext}} + V_{21}^{\text{b}}(1-G_{12}) \Delta n_{1+} \nonumber   \\
    && \; + v(1-G_{2+}) \Delta n_{2+}      \label{C5}\\
    V_{2-}^{\text{eff}} &=& V_{2-}^{\text{ext}} +  v(-G_{2-}) \Delta n_{2-} \label{C6} 
\end{eqnarray}
The sum equations represent the response to a density disturbance such as a test charge. In this case both the electrons and positive fermions contribute to screening. 
%in the equations are coupled (??).                                                                                         

The difference equations are proportional to the magnetic response. Since the electrons and positive fermions do not exchange, they do not have a magnetic interaction. The electrons and positive fermions respond independently to a magnetic field and the total response is simply the sum of the responses of the individual gases.

The response of the electrons to an effective potential is known from the uniform electron gas. Using $\Delta n_{2+} = -\Pi^{0}_2 V_{2+}^{\text{eff}}$ and defining $\varepsilon_{\text{et}} = 1+ v(1-G_{2+}) \Pi^{0}_{2} $, Eqs. \eqref{C5} \& \eqref{C6} become the general results,
\begin{eqnarray}
    V_{2+}^{\text{eff}} &=& \frac{V_{2+}^{\text{ext}} + V_{21}^{\text{b}}(1-2 G_{21}) \Delta n_{1+}}{\varepsilon_{\text{et}}} \label{C7}   \\
    V_{2-}^{\text{eff}} &=& \frac{V_{2-}^{\text{ext}}}{1- G_{2- }v \Pi^{0}_2} 
\end{eqnarray}
If all of the interactions are coulomb, this is easily understood. The external test charge is reduced by the induced charge in the positive fermion gas, and the resulting net charge is screened by the electron test charge dielectric function.

Inserting Eq. \eqref{C7} in Eq. \eqref{c3}, and using $G_{12} = G_{21}$ and $V_{12}^{\text{b}} = V_{21}^{\text{b}}$, yields after some algebra the additional general result,
\begin{eqnarray}
    V_{1+}^{\text{eff}} &=& V_{1+}^{\text{ext}} - \frac{V_{2+}^{\text{ext}} V_{12}^{\text{b}}(1-2G_{12})\Pi^{0}_2}{\varepsilon_{\text{et}}} \label{C9} \\
    && \; -\left( V_{11}^{\text{b}}(1-G_{1+}) + \frac{ V_{12}^{\text{b}} (1-2G_{12})^2 \Pi^{0}_2}{\varepsilon_{\text{et}}} \right) \Delta n_{1+} \nonumber
\end{eqnarray}
This can be rewritten as
\begin{eqnarray}
    \!\!\!\!\! V_{1+}^{\text{eff}} \!\! &=& \!\! \!  V_{1+}^{\text{ext}} \!- \! \frac{V_{2+}^{\text{ext}} V_{12}^{\text{b}}(1-2G_{12})\Pi^{0}_2}{\varepsilon_{\text{et}}} \! - \! \frac{\omega_q^2 \Delta n_{1+}}{\frac{\omega_{p1}^2}{v}} \, ,    \\
     \text{\small{where}} && \!\!\!\!\!\!\! \omega_{q}^2 = \frac{\omega_{p1}^2}{v} \left( V_{11}^{\text{b}}(1-G_{1+})  + \frac{V_{12}^{\text{b}}(1-2G_{12})^2 \Pi^{0}_2}{\varepsilon_{\text{et}}}\right)  \nonumber
\end{eqnarray}      
The plasma frequency of the positive fermions can be written in several equivalent ways \cite{ref11,ref16,ref23},
\begin{eqnarray}
    \omega_{p1}^{2} &=& \frac{4 \pi e^2 n  }{M} \; = \; m v_{F2}^2 \frac{ q_{T\!F}^2 }{3M} \; = \; c_{B\!S}^2 \qtf^2
\end{eqnarray}
where $n$ is the electron and positive fermion density, $m$ is the electron mass, $M$ is the positive fermion mass, $v_{F2}$ is the electron Fermi velocity, $q_{TF}$ is the Thomas-Fermi wave vector of the electrons, and $c_{BS}$ is the Bohm-Staver speed of sound. 

A finite density response of the positive fermions $\Delta n_{1+}$ is possible even in the absence of an external disturbance if $(\omega_q^2/ (\omega_{p1}^2/v) = 0$. This determines the dispersion relation of the acoustic plasmons/phonons.

Using the equality $1- 1/ \varepsilon_{\text{tt}} = v \Pi^{0}_2/ \varepsilon_{\text{et}}$, the dispersion relation can be rewritten and directly compared to Eq. (48) in Ref. \cite{ref10}, which demonstrates the effect of exchange and correlation of the positive fermions on the dispersion relation. Ref. \cite{ref10} assumed that $G_{1+}$ and $G_{12}$ both were zero which treats the ions as classical particles. Eq. \eqref{C9} is identical to Eq. (2.12) of Ref. \cite{ref23} if the bare interactions are the Coulomb interaction.

If the positive fermions were simply point particles with the coulomb interaction, the induced density would be related to the effective potential through the Lindhard function, $ \Delta n_{1+} = -\Pi^0_1 V_{1+}^{\text{eff}}$, and the coupled equations can be solved. It is convenient to define a dielectric function between a positive fermion and a test charge in analogy to the electron-test charge dielectric function $\varepsilon_{\text{ht}} = 1+ V_{12}^b (1-G_{1+})\Pi^0_1$. The results are
\begin{eqnarray}
    V_{11}^{b} &=& \frac{\varepsilon_{\text{et}} V_{1+}^{\text{ext}} - V_{12}^b \Pi^0_2 (1-2G_{12}) V_{2+}^{\text{ext}}}{\Delta}  \\
    V_{2+}^{\text{eff}} &=& \frac{\varepsilon_{\text{ht}} V_{2+}^{\text{ext}} - V_{12}^b \Pi^0_1 (1-2G_{12}) V_{1+}^{\text{ext}}}{\Delta}  \\
    \Delta &=& \varepsilon_{\text{et}} \varepsilon_{\text{ht}} -(V_{12}^b)^2 \Pi^0_1 \Pi^0_2 (1-2G_{12})^2
\end{eqnarray}
These are the results obtained in Ref. \cite{ref3}, differing only by a factor of two in the definition of the electron-positive fermion local field factor $G_{12}$.  It follows that
\begin{eqnarray}
    V_{1-}^{\text{eff}} &=& \frac{V_{1-}^{\text{ext}}}{1-G_{1-} V_{11}^b \Pi^0_1} \label{eq_c15}  \\
    V_{2-}^{\text{eff}} &=& \frac{V_{2-}^{\text{ext}}}{1-G_{2-}v \Pi^0_2} \label{eq_c16}
\end{eqnarray}
These equations are quite general and are discussed in Section III. Two simple observations should be noted. If the disturbance is a Coulomb interaction, then the external interactions of the electrons and positive fermions will have opposite signs. This means that $V_{1+}^{\text{ext}} = - V_{2+}^{\text{ext}}$ and $V_{1-}^{\text{ext}} = V_{2-}^{\text{ext}} = 0$. There is no magnetic response to a density (charge) disturbance in the paramagnetic system. Similarly if the disturbance is a magnetic field, the sum external potentials equal zero and the difference potential equals the external potential. There is no density response to a magnetic disturbance.

These equations can also be used to calculate the electron-electron, positive fermion-positive fermion, and electron-positive fermion interactions that are needed for superconductivity and transport calculations. As described in Appendix B, one specific electron or positive fermion is considered the external disturbance.  The equations above are used to calculate the average potentials felt by an electron and positive fermion in the Fermi sea. These potentials are needed to calculate the induced densities of electrons and positive fermions. The second specific electron or positive fermion interacts in an average mean field sense with the induced densities of electrons and positive fermions. However, the two specific electrons or positive fermions interact with each other via the bare Coulomb interaction. Exchange and correlation between the two specific electrons or positive fermions are included by using antisymmetric wave functions for the matrix elements between two parallel spin electrons, and going beyond first-order perturbation theory.

These interactions are presented in a subsequent paper. 

The specific local field factor for a spin-up electron used in this paper is
\begin{eqnarray}
    G_{2+} &=& \left( 1- \frac{\kappa_0}{\kappa} \right) \left( \frac{q}{q_{TF}} \right)^2  \;  ,
\end{eqnarray}
where the bulk modulus, $B=1/ \kappa$, is obtained by differentiating the sum of the electron and positive fermion energies. The compressibility of the noninteracting electron gas \cite{ref16} is $\kappa_0= (6.13/r_s)5 *1010 \, \text{ dynes}/\text{cm}^2$, or can also be obtained by differentiating only the kinetic energy portion of the total energy. The Thomas-Fermi screening wave vector of the electrons is given by $q_{TF}= 2.95/r_s^{0.5}$ inverse angstroms \cite{ref16}. In the metallic region, $\kappa_0/\kappa$ is roughly equal to $1-r_s/5.25$. 

The positive fermions have an effective $r_s$ that is $M/m$ times that of the electrons. One can see that the compressibility ratio of the positive fermions becomes negative while the compressibility ratio of the electrons is positive.  The true instability point is when the bulk modulus of the sum of the two is zero. The scaling relationship yields $\Pi_1^0 =M/m * \Pi_2^0$, and the local field factor $G_{1+}$ is exactly the same as for the electrons except that the $\kappa_0 / \kappa$ of the positive fermions is evaluated at $M/m$ times the $r_s$ of the electrons.

These definitions plus the Lindhard function \cite{ref16} are everything needed to produce the results of this paper.

\bibliography{reference}

\begin{thebibliography}{26}
\providecommand{\natexlab}[1]{#1}
\providecommand{\url}[1]{\texttt{#1}}
\expandafter\ifx\csname urlstyle\endcsname\relax
  \providecommand{\doi}[1]{doi: #1}\else
  \providecommand{\doi}{doi: \begingroup \urlstyle{rm}\Url}\fi

\bibitem[Pines(1956)]{ref1}
David Pines.
\newblock Electron interaction in solids.
\newblock \emph{Canadian Journal of Physics}, 34:\penalty0 1378, 1956.

\bibitem[Brinkman and Rice(1973)]{ref2}
WF~Brinkman and TM~Rice.
\newblock Electron-hole liquids in semiconductors.
\newblock \emph{Physical Review B}, 7\penalty0 (4):\penalty0 1508, 1973.

\bibitem[Vashishta et~al.(1974)Vashishta, Bhattacharyya, and Singwi]{ref3}
P~Vashishta, P~Bhattacharyya, and KS~Singwi.
\newblock Electron-hole liquid in many-band systems. i. ge and si under large uniaxial strain.
\newblock \emph{Physical Review B}, 10\penalty0 (12):\penalty0 5108, 1974.

\bibitem[Vignale and Singwi(1985{\natexlab{a}})]{ref4}
Giovanni Vignale and Kundan~S Singwi.
\newblock Possibility of superconductivity in the electron-hole liquid.
\newblock \emph{Physical Review B}, 31\penalty0 (5):\penalty0 2729, 1985{\natexlab{a}}.

\bibitem[Hammerberg and Ashcroft(1974)]{ref5}
J~Hammerberg and NW~Ashcroft.
\newblock Ground-state energies of simple metals.
\newblock \emph{Physical Review B}, 9\penalty0 (2):\penalty0 409, 1974.

\bibitem[Chakravarty and Ashcroft(1978)]{ref6}
Sudip Chakravarty and NW~Ashcroft.
\newblock Ground state of metallic hydrogen.
\newblock \emph{Physical Review B}, 18\penalty0 (9):\penalty0 4588, 1978.

\bibitem[Chakraborty et~al.(1983)Chakraborty, Kallio, Lantto, and Pietil{\"a}inen]{ref7}
Tapash Chakraborty, A~Kallio, LJ~Lantto, and P~Pietil{\"a}inen.
\newblock Structure of liquid metallic hydrogen as a two-component fermi fluid at t= 0.
\newblock \emph{Physical Review B}, 27\penalty0 (5):\penalty0 3061, 1983.

\bibitem[Ceperley and Alder(1987)]{ref8}
David~M Ceperley and Berni~J Alder.
\newblock Ground state of solid hydrogen at high pressures.
\newblock \emph{Physical Review B}, 36\penalty0 (4):\penalty0 2092, 1987.

\bibitem[Richardson and Ashcroft(1996)]{ref9}
CF~Richardson and NW~Ashcroft.
\newblock Superconductivity in electron liquids with and without intermediaries.
\newblock \emph{Physical Review B}, 54\penalty0 (2):\penalty0 R764, 1996.

\bibitem[Kukkonen and Overhauser(1979)]{ref10}
Carl~A Kukkonen and A.~W. Overhauser.
\newblock Electron-electron interaction in simple metals.
\newblock \emph{Physical Review B}, 20\penalty0 (2):\penalty0 550, 1979.

\bibitem[Giuliani and Vignale(2005)]{ref11}
Gabriele Giuliani and Giovanni Vignale.
\newblock \emph{Quantum Theory of the Electron Liquid}.
\newblock Cambridge University Press, 2005.

\bibitem[Han et~al.(2019)Han, Zhang, and Dai]{ref12}
Zhaoyu Han, Shiwei Zhang, and Xi~Dai.
\newblock Charge density waves in a quantum plasma.
\newblock \emph{Physical Review B}, 100\penalty0 (15):\penalty0 155132, 2019.

\bibitem[Ashcroft(1968)]{ref13}
Neil~W Ashcroft.
\newblock Metallic hydrogen: A high-temperature superconductor?
\newblock \emph{Physical Review Letters}, 21\penalty0 (26):\penalty0 1748, 1968.

\bibitem[Perdew et~al.(1990)Perdew, Tran, and Smith]{ref18}
John~P Perdew, Huy~Q Tran, and Elizabeth~D Smith.
\newblock Stabilized jellium: Structureless pseudopotential model for the cohesive and surface properties of metals.
\newblock \emph{Physical Review B}, 42\penalty0 (18):\penalty0 11627, 1990.

\bibitem[Perdew and Wang(1992)]{ref17}
John~P Perdew and Yue Wang.
\newblock Accurate and simple analytic representation of the electron-gas correlation energy.
\newblock \emph{Physical review B}, 45\penalty0 (23):\penalty0 13244, 1992.

\bibitem[Vashishta and Kalia(1982)]{ref20}
P~Vashishta and RK~Kalia.
\newblock Universal behavior of exchange-correlation energy in electron-hole liquid.
\newblock \emph{Physical Review B}, 25\penalty0 (10):\penalty0 6492, 1982.

\bibitem[Vignale(1984)]{ref23}
Giovanni Vignale.
\newblock \emph{Collective modes, effective interaction and superconductivity in the electron-hole liquid}.
\newblock Northwestern University, 1984.

\bibitem[Kaplan and Kukkonen(2023)]{ref14}
Aaron~D Kaplan and Carl~A Kukkonen.
\newblock Qmc-consistent static spin and density local field factors for the uniform electron gas.
\newblock \emph{Physical Review B}, 107\penalty0 (20):\penalty0 L201120, 2023.

\bibitem[Kukkonen and Wilkins(1979)]{ref15}
Carl~A Kukkonen and John~W Wilkins.
\newblock Electron-electron scattering in simple metals.
\newblock \emph{Physical Review B}, 19\penalty0 (12):\penalty0 6075, 1979.

\bibitem[Moldabekov et~al.(2022)Moldabekov, Vorberger, and Dornheim]{ref24}
Zhandos Moldabekov, Jan Vorberger, and Tobias Dornheim.
\newblock Density functional theory perspective on the nonlinear response of correlated electrons across temperature regimes.
\newblock \emph{Journal of Chemical Theory and Computation}, 18\penalty0 (5):\penalty0 2900--2912, 2022.

\bibitem[Kukkonen and Chen(2021)]{ref22}
Carl~A Kukkonen and Kun Chen.
\newblock Quantitative electron-electron interaction using local field factors from quantum monte carlo calculations.
\newblock \emph{Physical Review B}, 104\penalty0 (19):\penalty0 195142, 2021.

\bibitem[Nazarov and Silkin(2024)]{ref25}
Vladimir~U Nazarov and Vyacheslav~M Silkin.
\newblock Exchange kernel fxh (q, $\omega$) of electron liquid from the variational principle of mclachlan.
\newblock \emph{Physical Review B}, 110\penalty0 (10):\penalty0 104310, 2024.

\bibitem[Moroni et~al.(1995)Moroni, Ceperley, and Senatore]{ref26}
Saverio Moroni, David~M Ceperley, and Gaetano Senatore.
\newblock Static response and local field factor of the electron gas.
\newblock \emph{Physical review letters}, 75\penalty0 (4):\penalty0 689, 1995.

\bibitem[Ashcroft and Mermin(1976)]{ref16}
N.~W. Ashcroft and N.~D. Mermin.
\newblock \emph{Solid State Physics}.
\newblock Holt, Rinehart and Winston, New York London, 1976.

\bibitem[Tran and Perdew(2003)]{ref19}
Hoang~T Tran and John~P Perdew.
\newblock How metals bind: The deformable-jellium model with correlated electrons.
\newblock \emph{American Journal of Physics}, 71\penalty0 (10):\penalty0 1048--1061, 2003.

\bibitem[Vignale and Singwi(1985{\natexlab{b}})]{ref21}
G~Vignale and KS~Singwi.
\newblock Effective two-body interaction in coulomb fermi liquids.
\newblock \emph{Physical Review B}, 32\penalty0 (4):\penalty0 2156, 1985{\natexlab{b}}.

\end{thebibliography}
%\printbibliography
%\input{main.bbl}
\end{document}